\pdfoutput=1
\documentclass{elsart}

\usepackage{epsfig}
\usepackage{cite}

\usepackage{amsmath}  
\usepackage{amssymb}
\usepackage{euscript}

\newcommand{\Keu}{\EuScript{K}}

\newcounter{bla}

\begin{document}
\begin{frontmatter}

\begin{flushright}
\bf IFJPAN-IV-2008-7\\
\end{flushright}

\title{Markovian Monte Carlo program {\tt EvolFMC v.2} for solving 
       QCD evolution equations\thanksref{tok}}

\thanks[tok]{
  The project is partly supported by the EU grant MTKD-CT-2004-510126,
  realized in the partnership with the CERN Physics Department, by the
  Polish Ministry of Science and Information Society Technologies
  grant No.\ 620/E-77/6.PR UE/DIE 188/2005-2008,
  by the EU Marie Curie Research Training Network grant 
  under the contract No.\ MRTN-CT-2006-035505
  and by the Polish Ministry of Science and Higher Education grant 
  No.\ 153/6.PR UE/2007/7.
}

\author[a]{S.~Jadach}
\author[b]{W.~P\l{}aczek}
\author[a]{M.~Skrzypek}
\author[a]{P.~Stok\l{}osa}

\address[a]{Institute of Nuclear Physics IFJ-PAN,\\
  ul.\ Radzikowskiego 152, 31-342 Cracow, Poland}
\address[b]{Marian Smoluchowski Institute of Physics, Jagiellonian University,\\
   ul.\ Reymonta 4, 30-059 Cracow, Poland}

\begin{abstract}
We present the program {\tt EvolFMC v.2} that solves the evolution equations
in QCD for the parton momentum distributions by means of the Monte Carlo
technique based on the Markovian process. The program solves the DGLAP-type
evolution as well as modified-DGLAP ones. In both cases the evolution can be
performed in the LO or NLO approximation. The quarks are treated as massless.
The overall technical precision of the code has been established at $5\times
10^{-4}$. This way, for the first time ever, we demonstrate that with the Monte
Carlo method one can solve the evolution equations with precision comparable to
the other numerical methods. 

\begin{flushleft}
PACS: 12.38.-t, 12.38.Bx, 12.38.Cy
\end{flushleft}

\begin{keyword}
Monte Carlo, evolution equations, Markovian process, radiative corrections, QCD,
NLO, DGLAP, LHC, HERA, PDF
\end{keyword}

\end{abstract}

\end{frontmatter}


{\bf PROGRAM SUMMARY/NEW VERSION PROGRAM SUMMARY}

\begin{small}
\noindent
{\em Manuscript Title:}  The Markovian Monte Carlo program {\tt EvolFMC v.2} for
solving QCD evolution equations \\
{\em Authors:}           S.~Jadach, W.~P\l{}aczek, M.~Skrzypek, P.~Stok\l{}osa
\\
{\em Program Title:} {\tt EvolFMC v.2}                                         \\
{\em Journal Reference:}                                      \\
{\em Catalogue identifier:}                                   \\
{\em Licensing provisions:}      none                             \\
{\em Programming language:}      C++                             \\
{\em Computer:}    PC, Mac                                          \\
{\em Operating system:}  Linux, Mac OS X                             \\
{\em RAM:} less than 256 MB                                  \\
{\em Number of processors used:} 1                             \\
{\em Supplementary material:}                                 \\
{\em Keywords:} Monte Carlo, evolution equations, Markovian process, radiative
corrections, QCD, NLO, LHC, HERA, DGLAP, PDF
\\
{\em PACS:}  12.38.-t, 12.38.Bx, 12.38.Cy                      \\
{\em Classification:}                                         \\
{\em External routines/libraries:}  {\tt ROOT}                       \\
{\em Subprograms used:}                                       \\
{\em Nature of problem:}\\
Solution of the QCD evolution equations for the parton momentum distributions of
the DGLAP- and modified-DGLAP-type in
the LO and  NLO approximations.
   \\
{\em Solution method:}\\
Monte Carlo simulation of the Markovian process of a multiple emission of partons.
   \\
{\em Restrictions:}\\ 
(1) Limited to the case of massless partons.\\
(2) Implemented in the LO and NLO approximations only.\\
(3) Weighted events only.
   \\
{\em Unusual features:}\\
Modified-DGLAP evolutions included up to the NLO level.
   \\
{\em Additional comments:}\\
Technical precision established at $5\times 10^{-4}$. \\
The {\tt EvolFMC version 1} was described
in \cite{GolecBiernat:2006xw}, but the actual code was not published.
   \\
{\em Running time:}\\
For the $10^6$ events at 100 GeV:
DGLAP NLO: 27s; 
C'-type modified DGLAP NLO: 150s (MacBook Pro with Mac OS X v.10.5.5, 2.4 GHz
Intel Core 2 Duo, gcc 4.2.4, single thread); 

\end{small}

\newpage

\hspace{1pc}
{\bf LONG WRITE-UP}

\section{Introduction}
The evolution equations (EVEQs) for parton distribution functions (PDFs) and
parton momentum distributions are one of the most efficient tools in calculating
the radiative corrections in Quantum Chromodynamics (QCD) because they perform
resummation of certain types of corrections up to infinite order. The PDFs are
indispensable in any analysis of scattering processes which involve hadrons. The
 non-perturbative information on hadron structure, for the time being not
calculable, is extracted from the experimental data and then used to create the
PDFs at low energy scale. 
Among various types of evolution equations the most important is the family of
the DGLAP-type equations \cite{DGLAP}. Other widely used types of the EVEQs
include BFKL \cite{BFKL}, CCFM \cite{CCFM} or IREE \cite{IREE}. In this paper we
will discuss the DGLAP and modified-DGLAP types of EVEQs. However, a comment on
the relation to CCFM EVEQs will be made. 

There are various numerical methods of solving the DGLAP-type EVEQs: 
Mellin transforms \cite{Weinzierl:2002mv}, evolution on a finite grid
\cite{Miyama:1995bd,qcdnum16,Chuvakin:2001ge}, expansion in Laguerre polynomials
\cite{Coriano:1998wj,Toldra:2001yz}, expansion in Chebyshev polynomials
\cite{APCheb40,GolecBiernat:2007xv}, etc. The Monte Carlo (MC) methods differ
from the other numerical techniques because, in addition to providing the
inclusive parton distributions, they supply also the complete tree of parton
emissions during evolution. This allows one to construct the MC Parton Shower
programs which provide the actual four-momenta of emitted quarks and gluons -- a
necessary input for any realistic analysis which must include experimental
apparatus effects. One can find numerous implementations of the leading order
DGLAP evolution in the MC Parton Shower codes; let us quote just a few
examples: 
 PYTHIA~\cite{Sjostrand:2007gs,Sjostrand:2000wi},
 HERWIG~\cite{Gieseke:2006ga,Corcella:2000bw},
 ARIADNE~\cite{Lonnblad:1992tz},
 GR$@$PPA~\cite{Kurihara:2002ne,Tsuno:2006cu}.
So far the MC methods  {\em have not been considered} as a realistic alternative
to the other numerical methods of solving EVEQs due to low precision and long
time of computation.
The presented here MC code {\tt EvolFMC} is intended to fill-in this gap,
profiting from
the dramatic increase of the CPU power over two decades since NLO DGLAP evolution
was formulated and solved numerically for the first time.
It will be demonstrated in the following with the examples of numerical calculations
that {\tt EvolFMC} can solve the (modified) DGLAP-type EVEQs with high precision
($5\times 10^{-4}$ at least) within a reasonable CPU time.

The program {\tt EvolFMC} solves the EVEQs for the parton momentum
distributions by means of the MC simulation of the multiple emission of partons
in the cascade. The emission process is of the Markovian type, i.e.\ each
emission depends on the information from the previous emission only. The
algorithms are constructed on the basis of the Markovian process with simplified
emission kernels which retain only the leading singularities. The complete
kernels in the leading and next-to-leading approximation are then recovered by
the standard reweighting procedure.

The evolution is two-dimensional ($x$ and $t$) by construction. 
However, the azimuthal angle can always be added with a flat probability density
distribution.
Having identified the evolution time with certain kinematical variable,
one can then reconstruct the four-momenta of all emitted partons.
Such a procedure of reconstructing the four-momenta is, of course, exact only in
the LO approximation. In the NLO approximation the differential distributions
will be strictly speaking correct only in the "inclusive" sense of the overall
normalisation. This is, however, the common (and in fact the only available)
approach to the Parton Shower MCs.
It should be mentioned here that recently there have been a few attempts to
construct a true NLO Parton Shower algorithm with the help of "fully
unintegrated" or "exclusive" partonic functions
\cite{Collins:2007ph,Nagy:2007ty,Jadach:2009gm,Slawinska:2009gn}.
 
Apart from the standard DGLAP, two modified-DGLAP-type EVEQs can be solved by 
{\tt EvolFMC}. The modifications involve a change of the argument of the
coupling constant together with the introduction of a finite cut-off on its
minimal value. The program works in the weighted mode
only. The first version of the code, the {\tt EvolFMC v.1}, was described
in \cite{GolecBiernat:2006xw}, but the actual code was not published. The
version {\tt v.1} solved modified-DGLAP-type in the LO approximation only. The
NLO evolution was available only in the standard DGLAP case. Also the structure
of the code is rebuilt in version {\tt v.2}.

Let us conclude this introduction with the following remark.
The MC approach to EVEQs has one important general advantage: with the help of
the reweighting technique it is easy to introduce into evolution some additional
effects or modifications. As an example let us mention the possibility of
emulating the CCFM-type evolution. From the algorithmic point of view, having
generated the azimuthal angles and reconstructing the transverse momenta of the
partons in the shower, one can trivially construct the so-called ``non-Sudakov''
form factor and include it as a correcting weight. Of course, one has
to remember that from the theoretical point of view the definition of
the non-Sudakov form factor in the NLO case is a highly nontrivial
problem; even the complete LO analysis is difficult in the context of
the CCFM equation \cite{Andersson:2002cf}. Additional modifications in evolution
kernels can be done as well.

The paper is organized as follows. In Section~\ref{sect:general} we give a short
theoretical overview of the evolution equations and their solutions by means of
MC methods. Section~\ref{sect:software} describes in some detail the architecture
of the presented MC code, {\tt EvolFMC version~2}. Section~\ref{sect:install}
contains instructions on how to install the code on {\tt Linux} and {\tt Mac
OS X} platforms. In Section~\ref{sect:demos} we present two demonstration programs
included in the distributed version of the code. Section~\ref{sect:precision}
contains a detailed study of the technical precision of {\tt EvolFMC}
at the level of $5\times 10^{-4}$. A short summary in Section~\ref{sect:summary} 
concludes the paper.

\section{Theoretical background}
\label{sect:general}
In this short theoretical overview we will only give a few formulae and
definitions  necessary to explain the notation, to define the problem and to
present the solution. For detailed derivations and technical description of the
algorithms we refer the reader to the extensive bibliography, which we will
present here in more detail. The short and elementary description of the
solution of the DGLAP-type EVEQs in terms of the Markovian process and its MC
realization in the LO approximation has been presented in Ref.\
\cite{Jadach:2003bu}.
 The extension to the NLO level and comprehensive description of MC algorithms
that solve the Markovian evolutions of PDFs and parton momentum distributions
has been given in Refs.\ \cite{GolecBiernat:2006xw,Placzek:2007xb}. The
algorithms presented in \cite{GolecBiernat:2006xw} have been implemented in the
first version of the code, {\tt EvolFMC v.1}.
 The extension to the modified-DGLAP-type evolutions and the detailed
description of appropriate modified-DGLAP Markovian algorithms has been given in
Refs.\ \cite{GolecBiernat:2007pu,Skrzypek:2007zz} for the LO case (implemented
in {\tt EvolFMC v.1}) and in Ref.\ \cite{Stoklosa:2008nj} for the NLO
approximation. The implementation of algorithms from Ref.\
\cite{Stoklosa:2008nj} as well as re-organization of the implementation of the
DGLAP-type algorithms from Ref.\ \cite{GolecBiernat:2006xw} has been done in the
second version of the code, {\tt EvolFMC v.2}, presented in this paper.
Finally, in Ref.\ \cite{GolecBiernat:2007xv} the general and universal formalism 
of constructing Markovian algorithms, common for all DGLAP-type and
modified-DGLAP-type evolutions, has been presented on the basis of the operator
language. It is this formulation \cite{GolecBiernat:2007xv} that we will use in
the rest of this section to describe the principles of the Markovian MC
evolution.

The evolution equation and its solution in the form of a master iterative
formula for the Markovian MC algorithm can be expressed as follows:
\begin{equation}
\label{eq:op-evoleq}
  \partial_t {\bf D}(t) = {\bf K}(t) \; {\bf D}(t),
\;\;\;\hbox{i.e. }\;\;
  \partial_t D_f(t,x)
 = \sum_{f'} \int_0^1 dw\; \Keu_{ff'}(t,x,w) D_{f'}(t,w),
\end{equation}
and
\begin{equation}
\label{eq:evol5}
\begin{split}
\bar{\bf E}{\bf D}(t)=
\int_{t_{0}}^{t} dt_1    \bigg(
\int_{t_{1}}^{t} dt_2    \bigg[&
\int_{t_{2}}^{t} dt_3    \bigg\{ \dots
\\
\dots
\int_{t_{N-1}}^{t} dt_{N}\bigg\{
&  \bar{\bf E}{\bf K}^R(t_{N})
   {\bf G}_{{\bf K}^V}(t_{N},t_{N-1})
  +\bar{\bf E}{\bf G}_{{\bf K}^V}(t,t_{N-1})\delta_{t_{N}=t}
\bigg\} \times
\\~~~~~~~~~~~~~ \vdots
\\ \times
&  {\bf K}^R(t_{2})
   {\bf G}_{{\bf K}^V}(t_{2},t_{1})
  +\bar{\bf E}{\bf G}_{{\bf K}^V}(t,t_{1})\delta_{t_2=t}
\bigg]\times
\\ \times
&
 {\bf K}^R(t_{1})
 {\bf G}_{{\bf K}^V}(t_{1},t_{0})
+\bar{\bf E}{\bf G}_{{\bf K}^V}(t,t_{0})\delta_{t_1=t}
\bigg)
{\bf D}(t_0).
\end{split}
\end{equation}
Let us describe all the ingredients of eqs.\ (\ref{eq:op-evoleq}) and (\ref{eq:evol5}).
\begin{itemize}
\item
The multiplication of the matrices is understood as: $\sum_{f'} \int_0^1 dw$.
\item
$D_{f}(t,w)$ is the parton density function of the parton $f$.
\item
$\Keu_{ff'}(t,x,w)$ is the generalized evolution kernel built from the real and virtual
parts:
\begin{align}
\label{eq:evkernel}
\begin{split}
\Keu_{ff'}(t,x,w)=& \Keu^V_{ff'}(t,x,w)+\Keu^R_{ff'}(t,x,w),
\\
 \Keu^V_{ff'}(t,x,w)=& -\delta_{ff'}\delta_{x=w}\Keu^v_{ff}(t,x).
\end{split}
\end{align}
\item
The operator $\bar{\bf E}$ is defined as $\{\bar {\bf E}\}_f(x)\equiv x$,
i.e.\ it turns parton distributions into parton momentum distributions, whereas
summing and integrating over final degrees of freedom means in the MC language that
we generate all possible final state configurations
without any constraints.
\item
${\bf G}_{{\bf K}^V}$ is the solution of the evolution
equation with the virtual kernel $\Keu^V_{ff'}(t,x,w)$ only
\begin{equation}
\label{eq:phidef}
\{{\bf G}_{{\bf K}^V}(t,t')\}_{ff'}(x,w)
  =\delta_{ff'}\delta_{x=w}\; e^{-\Phi_f(t,t'|w)}.
\end{equation}
\item
$\Phi_f(t,t'|x)$ is the Sudakov form factor, expressed in terms of the real
emission part of the evolution kernel
\begin{align}
\Phi_f(t,t'|x)=&\int\limits_{t'}^{t} dt''\; \Keu^v_{ff}(t'',x)
  =\sum_{f'} 
   \int\limits^{t}_{t'} d t'' 
   \int\limits_0^{x} \frac{d x'}{x} x'\;
      \Keu^R_{f' f}(t'',x',x)
\notag
\\
  =&\sum_{f'} \Phi_{f' f} (t, t' |x).
\label{eq:phireal}
\end{align}
\end{itemize}
The actual, normalized to unity, probability densities of the variables in each
step of the Markovian process are visible in each of the lines of the eq.\
(\ref{eq:evol5}), representing a single step in the emission chain:
\begin{equation}
\label{eq:omega}
\begin{split}
1&= \frac{1}{x_{i-1}}
\int\limits_{t_{i-1}}^{t} dt_i\;
\Big\{
  \bar{\bf E}{\bf K}^R(t_{i})
   {\bf G}_{{\bf K}^V}(t_{i},t_{i-1})
  +\bar{\bf E}{\bf G}_{{\bf K}^V}(t,t_{i-1})\delta_{t_{i}=t}
\Big\}_{f_{i-1}}(x_{i-1})
\\
&=
  e^{-\Phi_{f_{i-1}}(t,t_{i-1}|x_{i-1})}
+\int\limits^1_{e^{-\Phi_{f_{i-1}}(t,t_{i-1}|x_{i-1})}}
 d\left(e^{-\Phi_{f_{i-1}}(t_{i},t_{i-1}|x_{i-1})}\right)
\\ & ~~~~\times
 \biggl[
 \sum_{f_i}
 \frac{\partial_{t_i}\Phi_{f_i f_{i-1}}(t_{i},t_{i-1}|x_{i-1})}%
      {\partial_{t_i}\Phi_{f_{i-1}}(t_{i},t_{i-1}|x_{i-1})}
\\ & ~~~~~~~~\times
 \int dx_i\;
 \frac{1}{\partial_{t_i}\Phi_{f_i f_{i-1}}(t_{i},t_{i-1}|x_{i-1})}
 \frac{x_i}{x_{i-1}}
 \Keu^R_{f_if_{i-1}}(t_{i},x_i,x_{i-1})
 \biggr].
\end{split}
\end{equation}
In the program {\tt EvolFMC v.2} we have implemented three types of the 
evolution differing by the definition of the evolution kernel
$\Keu^{R(X)}_{f'f}(t,x,w)$, $X=A,B',C'$:
\begin{equation}
\label{kernBare}
\begin{split}
x\Keu^{R(X)}_{f'f}(t,x,w)
  =&   \theta_{t+\ln\phi_X>\ln\lambda}   \biggl[
  \frac{\alpha_{NLO}(t+\ln\phi_X)}{2\pi}2zP_{f'f}^{R(0)}(z)
\\&
+ \Bigl(\frac{\alpha_{NLO}(t+\ln\phi_X)}{2\pi}\Bigr)^2 2z
    \Bigl(P_{f'f}^{R(1)}(z)+\Delta P_{f'f}^{R(1)X}(z)\Bigr)
 \biggr],
\end{split}
\end{equation}
where $z=x/w$. The parameter $\lambda$ is an arbitrary cut-off on the  argument
of the coupling constant, greater than $\Lambda_{QCD}$, necessary in order to
avoid the singularity in the coupling constant. Note that the part of the real
emission phase space excluded by the cut-off $\lambda$ is compensated for by the
virtual form factor defined in eq.\ (\ref{eq:phireal}) as the integral over
phase space of a real emission. As a consequence the momentum sum rule is
preserved.
$\ln\phi_X$ takes one of the following three forms:
\\
A\phantom{'}: $\ln\phi_X=0$,~~ (the DGLAP evolution)
\\
B': $\ln\phi_X=\ln(1-z)$,~~ (the modified-DGLAP B'-type evolution)
\\
C': $\ln\phi_X=\ln\bigl(w(1-z)\bigr)$,~~ (the modified-DGLAP C'-type evolution).

The term $\Delta P_{f'f}^{R(1)X}(z)$ is added to remove the double counting
caused by the change of the argument of the coupling constant, according to the
prescription of Ref.\ \cite{Roberts:1999gb}. Namely, from the expansion 
$$\alpha_{NLO}(t+\ln\phi) =
 \alpha_{NLO}(t) -(\beta_0/2\pi)\alpha_{NLO}^2(t) \ln\phi
 +{\cal O}(1/t^3)$$ one obtains
\begin{align}
\label{ctrterm}
&\Delta P_{f'f}^{R(1)B'}(z) =
\Delta P_{f'f}^{R(1)C'}(z) =
  {\beta_0} \ln(1-z) P_{f'f}^{R(0)}(z).
\end{align}
Note that in the case C' we use the counter term identical as in the B' case. 
The additional piece related to $\ln w$ is of a genuine beyond-DGLAP origin,
i.e.\ it is absent in the DGLAP kernel. Therefore, there is no double counting
and no need to subtract it.
The universal LO part $P_{f'f}^{R(0)}(z)$ and the NLO part $P_{f'f}^{R(1)}(z)$ are given in 
\cite{Curci:1980uw,Furmanski:1980cm}.
Finally, the coupling constant at the NLO level has the standard
form
\begin{align}
\label{alpha}
    \alpha_{LO}(t) = \frac{2\pi}{\beta_0(t-\ln\Lambda_0)},
\;\;
    \alpha_{NLO}(t) = \alpha_{LO}(t)\left(
          1-\alpha_{LO}(t) \
              \frac{\beta_1 \ln(2t-2\ln\Lambda_0)}{4\pi\beta_0}
                                        \right).
\end{align}
On the technical side, in the actual MC algorithms implemented in {\tt EvolFMC
v.2} we do not use the complicated kernels (\ref{kernBare}). Instead, a series
of simplified kernels $\bar{\Keu}^{R(X)}$ is introduced. Each of them is chosen
in such a way that it retains only the leading singularities of the   exact
kernel ${\Keu}^{R(X)}$ while all the complicated but finite structure is
temporarily discarded:
\begin{align}
\label{simplea}
x\bar\Keu^{R(A)}_{f'f}(t,x,w)
\equiv&
\frac{\alpha_{LO}(t)}{2\pi}
2zP^{R(0)}_{f'f}(z),
\end{align}
for the $X=A$ case and
\begin{align}
\label{simplebprim}
x\bar\Keu^{R(X)}_{f'f}(t,x,w)
\equiv&
\frac{\alpha_{NLO}(t+\ln\phi_X)}{2\pi}
2z\bar P^{R(0)}_{f'f}(z)\theta_{t+\ln\phi_X>\ln\lambda},
\\
\notag
z\bar P^{R(0)}_{f'f}(z)=&
\frac{1}{1-z}
    (\delta_{f'f} A_{ff}^{(0)}+\max_z F^{(0)}_{f'f}(z) +M_{f'f}^{NLO})
\end{align}
for the cases $X=B'$ and $X=C'$.
The variable $z=x/w$.
The constant $M_{f'f}^{NLO}$ is defined as
\begin{align}
M_{f'f}^{NLO} = \left\{ \begin{array}{c}
0,\;\;\;\;{\rm if\;} P^{R(0)}_{f'f}(z)\neq 0,
\\
\eta,\;\;\;\;{\rm if\;} P^{R(0)}_{f'f}(z) = 0,
\end{array} \right.
\end{align}
and $\eta$ is a dummy technical parameter. The functions $A_{ff}^{(0)}(z)$ and
$F_{f'f}^{(0)}(z)$ are a convenient parametrization of the full LO kernels
$P_{f'f}^{R(0)}(z)$
\begin{align}
zP_{f'f}^{R(0)}(z) =&\frac{1}{1-z}\delta_{f'f} A_{ff}^{(0)}
                    +F_{f'f}^{(0)}(z),
\end{align}
see Appendix C of Ref.\ \cite{GolecBiernat:2006xw} for the complete list of
them (for example, for the $zP_{qq}^{R(0)}(z)$ kernel we have
$A_{qq}^{(0)}=2C_F$ and $F_{qq}^{(0)}(z)=C_F(-2-z-z^2)$). The NLO kernels
$zP_{qq}^{R(1)}(z)$ in the form used in the code are explicitly given in
Appendix A of Ref.\ \cite{GolecBiernat:2006xw}.
The exact kernel ${\Keu}^{R(X)}$ is recovered at the end by the standard
reweighting procedure. The correcting weight is 
\begin{align}
\label{wtcorr}
w^{(n)}
=& e^{\bar\Phi_{f_{n}}(t,t_{n} |x_{n})-\Phi_{f_{n}}(t,t_{n} |x_{n})}
\notag
\\ &\times
  \left(
  \prod_{i=1}^n 
\frac{      \Keu^R_{f_if_{i-1}}(t_{i},x_i,x_{i-1})}%
           {\bar{\Keu}^R_{f_if_{i-1}}(t_{i},x_i,x_{i-1})} 
  e^{\bar\Phi_{f_{i-1}}(t_{i},t_{i-1} |x_{i-1})
          -\Phi_{f_{i-1}}(t_{i},t_{i-1}|x_{i-1})}
  \right).
\end{align}
The product runs over all generated partons in a given MC event with multiplicity $n$, and the 
form factor $\bar\Phi_{f_{i-1}}(t_{i},t_{i-1}|x_{i-1})$ is constructed from
$\bar{\Keu}^R_{f_if_{i-1}}(t_{i},x_i,x_{i-1})$, in analogy to eq.\
(\ref{eq:phireal}). 

The actual expressions for the form factors $\Phi$ and $\bar\Phi$ are fairly
complicated, especially for the cases B' and C', and we will not quote them
here, referring the interested reader to the original papers. Let us only remark
that, as seen in eq.\  (\ref{eq:phireal}), the form factors are defined as
two-dimensional integrals. In the case of the simplified form factor $\bar\Phi$
both integrals can be done analytically%
\footnote{%
In fact this analytical integrability is one of the criteria in choosing the
form of the simplified kernels (\ref{simplebprim}). This is for example why the
$1/(1-z)$ term multiplies artificially also the $F$-function in eq.\
(\ref{simplebprim}). The constant $M$ is added to avoid potentially dangerous
zeroes at the NLO level.
}.
On the contrary, in the full form factor $\Phi$ only one integration
can be done analytically. The other one has to be done numerically on 
an event-per-event basis.

\section{Overview of the software structure}
\label{sect:software}
The program {\tt EvolFMC} is written in the {\tt C++} language.
To compile and link the code we use
the {\tt autotools} utility.
It  allows us to compile/link the code on many platforms in a simple way.
The code has been routinely compiled and run
under the {\tt Linux} and {\tt Mac~OS~X 10.5} operating systems.
From the user point of view, the only difference between these two systems is in the compiler's options inside the file {\tt configure.in}, which is included in the main
folder of the project.
The central part of the {\tt EvolFMC} source code is the {\tt MarkovMC}
library located in the {\tt MarkovMC} folder.
This folder includes the essential source code necessary to solve evolution equations.
This part of the code requires only basic {\tt C++} libraries and an external
random number generator (RNG).
We use the generator {\tt TRandom3} from the {\tt ROOT}
package as a default RNG.
With this generator we have reached the precision below $0.05\%$.
In case when {\tt ROOT} is not available on a given system platform,
one should replace in the {\em wrapper} class {\tt rndm}
the name {\tt TRandom3} with the name of the other RNG.
A simple main program in the {\tt Demo0} folder uses only
the standard {\tt C++} libraries and can be built and executed without {\tt ROOT}.
On the other hand, the more sophisticated source code in the {\tt Demo1} folder
of the distribution version of the project uses the {\tt ROOT} library,
mainly for booking, filling and drawing histograms and more, see below.

The library in the {\tt MarkovMC} folder has a modular structure
in the sense that the algorithms that solve a particular type of the evolution equations
are implemented in separate classes and located in separate source files.
Each class implementing the Markovian MC algorithm for a given type of evolution equation
includes all formulae, in particular the Sudakov form factors, specific to a given
evolution type.
All classes specific to one type of evolution
inherit from the common base class {\tt MarkovianGen}.
This modular structure also makes it easier to add in the future
any new type of the QCD evolution of the parton distributions to the code.

\subsection{Structure of folders}
In the following the structure of the folders of {\tt EvolFMC} is described:
\begin{itemize}
\item  {\tt MarkovMC}
-- the folder containing the library of the Markovian MC engines.
Source codes of the classes solving various types of the evolution equations
are placed in separate files.
\item {\tt Demo0}
-- the folder with the simple demonstration program {\tt Demo0}
written in the {\tt C} language.
This program demonstrates the standalone usage of the library {\tt MarkovMC},
without the use of {\tt ROOT}.
\item  {\tt Demo1}
--  the folder hosting the demonstration program {\tt Demo1},
a template program for the advanced user of {\tt EvolFMC}.
{\tt Demo1} requires {\tt ROOT} to be installed in the system.
The subfolder {\tt Demo1/work} contains
scripts necessary to run {\tt Demo1}.
\item {\tt m4}
-- the folder containing a script which defines properly a path for the
{\tt ROOT} libraries.
This script is used by the {\tt automake} program.
\end{itemize}
Each of the above folders contains also scripts (in the files {\tt Makefile.am}),
which are required by the {\tt automake} utility.

\subsection{Source code in folder MarkovMC}
The source code of the {\tt MarkovMC} library consists of three categories
 of files:
(1) the files containing base classes of the Markovian MC generator,
(2) the files which contain classes specific to a particular
    type of the QCD evolution equations, and
(3) the files with some auxiliary classes.
\begin{enumerate}
\item { Base classes}
\begin {itemize}
\item {\tt markoviangen.cxx, markoviangen.h}:\\
  The class {\tt MarkovianGen} contains the essential part of the Markovian
  MC algorithm,
  member functions and data members common to all types of the QCD evolution.
  In particular, it executes the main Markovian loop over parton emissions.
  Virtual member functions encapsulate evolution details.
\item {\tt kernels.cxx, kernels.h}:
  This class defines the DGLAP LO kernels.
\item {\tt kernels\_nlo.cxx, kernels\_nlo.h}:
  This class defines the DGLAP NLO kernels;
  it inherits from the simpler class {\tt kernels}.
\end {itemize}
\item { Auxiliary classes}
   \begin {itemize}
   \item {\tt gaussintegral.cxx, gaussintegral.h}:
      the standard Gauss integration procedure,
      translated from the {\tt Fortran  GNU} library to {\tt C++}.
   \item {\tt rndm.cxx, rndm.h}:
      the wrapper class of the random  number generator;
      it is derived from the {\tt ROOT} class {\tt TRandom3}.
   \end {itemize}
\item {Classes implementing one particular type of the QCD evolution equation
for the parton momentum distributions:}
  \begin {itemize}
  \item {\tt dglap\_lo.cxx, dglap\_lo.h}:
     implementation of the DGLAP LO evolution;
     this class is derived from the classes {\tt MarkovianGen} and {\tt kernels}.
  \item {\tt dglap\_nlo.cxx dglap\_nlo.h}:
     implementation of the DGLAP NLO evolution;
     this class is derived from the classes {\tt MarkovianGen} and {\tt kernels\_nlo}.
  \item {\tt bprim\_lo.cxx, bprim\_lo.h}:
     implementation of the modified-DGLAP evolution scheme B', the LO case;
     this class is derived from the classes {\tt MarkovianGen, kernels}
     and {\tt gaussIntegral}.
  \item {\tt bprim\_nlo.cxx, bprim\_nlo.h}:
     implementation of the modified-DGLAP evolution scheme B',
     the NLO case, the basic algorithm;
     this class is derived from the classes {\tt MarkovianGen, kernels\_nlo} and
     {\tt gaussIntegral}.
  \item {\tt bprim\_nlo\_aux.cxx, bprim\_nlo\_aux.h}:
     implementation of the modified-DGLAP evolution scheme B', the NLO case,
     the auxiliary algorithm (for tests only!);
     this class is derived from the classes
     {\tt MarkovianGen, kernels\_nlo} and {\tt gausIntegral}.
  \item {\tt cprim\_lo.cxx, cprim\_lo.h}:
     implementation of the modified-DGLAP evolution scheme C', the LO case;
     this class is derived from the classes {\tt MarkovianGen, kernels}
     and {\tt gaussIntegral}.
  \item {\tt cprim\_nlo.cxx, cprim\_nlo.h}:
     implementation of the modified-DGLAP evolution scheme C',
     the NLO case, the basic algorithm;
     this class is derived from
     the classes {\tt MarkovianGen, kernels\_nlo} and {\tt GaussIntegral}.
  \item {\tt cprim\_nlo\_aux.cxx, cprim\_nlo\_aux.h}:
     implementation of the modified-DGLAP evolution scheme C', the NLO case,
     the auxiliary algorithm (for tests only!);
     this class is derived from the classes {\tt MarkovianGen, kernels\_nlo}
     and {\tt gaussIntegral}.
  \end {itemize}
\end{enumerate}

\subsection{Inheritance pattern of classes of {\tt MarkovMC} library.}

As already indicated,
all the classes that are used to solve the evolution equations for the parton
momentum distributions
by means of the Markovian MC method are derived from a few base classes.
Depending on the type of the QCD evolution, the base classes are:
{\tt MarkovianGen}, {\tt kernels} and its derived class {\tt kernels\_nlo}, and
the auxiliary class {\tt gaussIntegral}.
The derived classes implement details of the particular
type of the QCD evolution equations (e.g. {\tt dglap\_lo.cxx}).
In particular, these classes include member functions which generate
randomly the evolution time, the parton flavor and the $z$-variable,
which are defined as virtual member functions in the base class {\tt MarkovianGen}.
The base class {\tt MarkovianGen} implements all essential parts of
the Markovian algorithm and a few auxiliary functions.
The central member function of this class is {\tt GenerateEvent}.
The class {\tt MarkovianGen} does not know the details of the evolution kernels
-- they are implemented in the class {\tt kernels} and/or {\tt kernels\_nlo}.

The {\tt gaussIntegral} class owns integration methods. These methods are
necessary to calculate the Sudakov form factors for more complicated
types of the QCD evolution.

\begin{figure}[!ht]
  \centering
    \epsfig{file=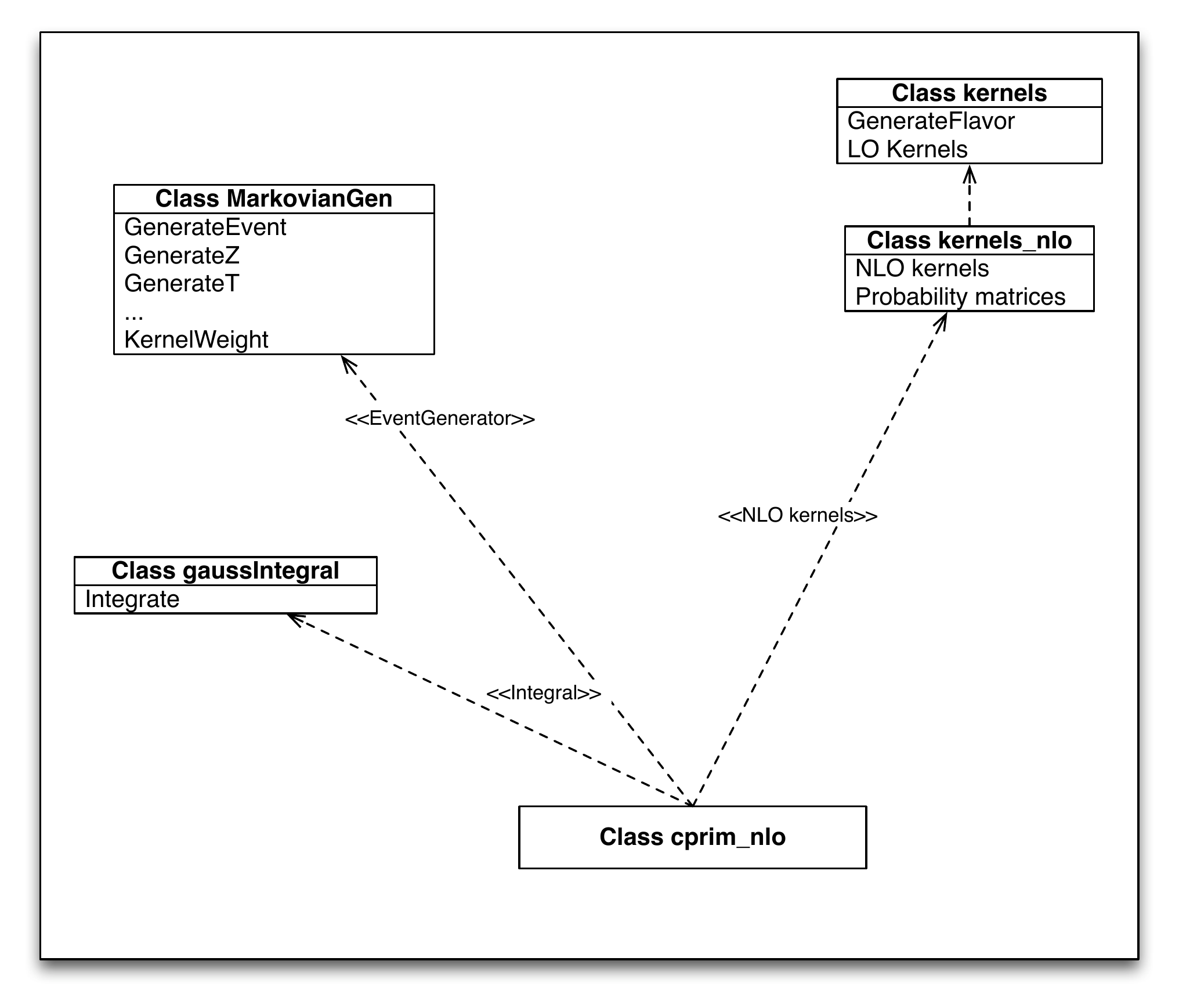,width=120mm}
    \caption{The structure of the derivations in the case of
             the modified-DGLAP NLO C'-type evolution.
    }
  \label{fig:ccfmnlouml}
\end{figure}

Depending on the complexity of the equation, the structure of inheritance has different forms.
As an example we present in Fig. \ref{fig:ccfmnlouml} the inheritance scheme for
the most complicated case of the modified-DGLAP NLO C'-type evolution.
\subsection{General design of {\tt MarkovMC} library}

The classes of the {\tt MarkovMC} library and the organization of the source
code
have been designed in such a way that
(a) an infrastructure is available for any kind of the Markovian
MC implementing
any kind of the QCD evolution,
(b) it is easy to include or exclude any group of classes implementing any
type of the QCD evolution.
It is therefore not surprising that
the classes which solve different evolution types are completely independent from each other.
One can exclude a particular class from the code without loosing functionality of other classes.
Also, adding more evolution types would not modify the structure of the library.
In the practical application, if one needs, for instance,
a solution of the modified-DGLAP NLO scheme C', then one should include only the header file
{\tt cprim\_nlo.h}.
A simple example will be shown in the {\tt Demo0} program in the following sections.

\subsection{Description of base class {\tt MarkovianGen}}

{\bf Virtual member functions.}  \\
The member functions from this group are implemented in the derived classes:
\begin{itemize}
\item
{\tt void MarkovianGen::GenerateEvent(double \&t, double \&Vx,
               double \&weight, double tmax, int \&Flavor,
                                 double epsTSolver = 0.0001)}
-- generates a single MC event according to the Markovian algorithm.
\item
{\tt void MarkovianGen::GenerateEvent()}
-- generates a single MC event according to the Markovian algorithm,
a ``wrapper'' function, see Sect.\ \ref{Demo1} for explanation.
\item
{\tt double GenerateT(double rndm, double t\_prev, double TStop, double epsTSolver, double Vx)}
-- generates the evolution time%
\footnote{
 The technical parameter {\tt epsTSolver} sets a precision for the {\tt TSolver} function
 used for some evolution types.}.
\item
{\tt double GenerateZ(double rndm, double t, double T0, double epsTSolver, double Vx)} --
generates the actual light-cone variable {\tt z}.
\item
{\tt int GenerateFlavor(double rndm,int oldFlavor)} --
    generates the parton-flavor index.
\item
{\tt double KernelWeight(double t, double z, double Vx) } --
provides part of the MC weight turning the simplified kernel into the exact kernel.
\item
{\tt double DeltaRealPart(double t\_new, double t\_old, double z)} --
part of the MC weight from the Sudakov form factor evaluated analytically.
\item
{\tt double DeltaVirtualPart(double t\_new, double t\_old, double z)} --
part of the MC weight due to the Sudakov form factor evaluated numerically.
\item
{\tt virtual void init()} --
initialization of an object.
\item {\tt void AddParticle(double t, double x, double z, int f, double weight,
int index)} -- 
stores data of a single parton emission, a private function, 
not to be used outside the {\tt MarkovianGen} class.
\item
Other auxiliary member functions used to transmit the flavor type
between the class {\tt MarkovianGen} and the class {\tt kernels}:
\begin{itemize}
    \item {\tt void SetActualFlavor(int flavor)} --
    transmits the actual parton flavor to {\tt kernels}.
    \item {\tt void SetOldFlavor(int flavor)} --
    transmits the old flavor to {\tt kernels}.
\end{itemize}
\item
``Getter'' functions for the external users:
\begin{itemize}
\item {\tt bool GetParticle(double \&t, double \&x, double \&z, int \&f, double
\&weight, int index)}
 -- provides the external user with the information about partons generated in the
last MC event; {\tt index} runs from $0$ to {\tt EventMultiplicity}, the variables 
{\tt t}, {\tt x}, {\tt z}, {\tt f} and {\tt weight} describe the emission of the number
{\tt index}, {\tt weight} is the cumulative weight; {\tt index}~$ = 0$ returns
information on the initial parameters of {\tt GenerateEvent} and in this case
{\tt z} $=1$. 
 \item {\tt int GetEventMultiplicity()} -- returns a number of particles generated
in the last MC event. 
\end{itemize}
\end{itemize}
All parameters of the virtual functions belong to the following list:
\begin{itemize}
	\item {\tt double rndm} -- the random number,
	\item {\tt double t } -- the evolution time,
	\item {\tt double T0} -- the start of the evolution time,
	\item {\tt double TStop} -- the end of the evolution time,
	\item {\tt double z} -- $z={x_{\rm new}}/{x_{\rm old}}$,
	\item {\tt double Vx} -- as input: the light-cone $x_{\rm old}$ variable
before the emission; as output: the $x_{\rm new}$ variable after the emission,
	\item {\tt double t\_new} -- $t_{\rm new}$ the generated current
evolution time,
	\item {\tt double t\_old} -- $t_{\rm old}$ the evolution time of
the previous emission,
	\item {\tt double knew} -- $f_{\rm new}$ the generated current flavor,
	\item {\tt double kold} -- $f_{\rm old}$ the flavor of the previous
emission,
	\item {\tt double flavor} -- the flavor (depends on the function -- explanation above),
	\item {\tt double weight} -- as input: the initial weight to be
assigned to the event (for example from the generation of the initial
condition); 
as output: the cumulative weight of the event,
	\item {\tt epsTSolver} -- the precision for the {\tt TSolver} function.
\end{itemize}
\subsection{Small parameters in classes of {\tt MarkovMC} library}
\label{smallparams}
Constructors and other methods of the classes in the {\tt MarkovMC} library have
as formal parameters several small parameters, which we call epsilon-parameters.
They may be of technical or physical character.
Let us explain them with explicit examples.

(1) The constructor
\begin{verbatim}
dglap_nlo(double T0, double lambda, double NumberOfFlavor,
                 rndm *rnGen, double epsilon_IRC = 0.0001,
                              double epsilon_Zmin = 0.0001)
\end{verbatim}
contains two physical epsilon-parameters:
\begin{itemize}
\item
  {\tt epsilon\_IRC} is the {\em dummy} infrared cut-off at $z=1$ for the DGLAP
evolution. Solutions of the evolution equations do not depend on its value as
long as it is kept small enough ($10^{-4}$ by default).
\item
  {\tt epsilon\_Zmin} is the minimal value of the final $x$-variable. 
  In practice it is used as a cut-off for the $z$-variable in the kernels
  which exhibit a logarithmic divergency in the small $z$ limit.
  Note that the MC program will generate the distribution for $x<\epsilon_{\rm Zmin}$
  but it will be incorrect, see Sect.\ \ref{simpledemo} for more comments.
\end{itemize}

(2) In more advanced classes, such as {\tt cprim\_nlo}, there is another
technical epsilon-parameter: {\tt epsTSolver}.
It is used in the method {\tt GenerateEvent} of the base class to
set the precision of the
important {\tt TSolver} member function which inverts numerically an arbitrary
one-dimensional function.

(3) The last two technical epsilon-parameters are defined in
the auxiliary class {\tt gaussIntegral},
see for instance one of its member functions
\begin{verbatim}
double dqags(double a, double b, double epsabs, double epsrel,
             double &abserr, int &neval, int &ier),
\end{verbatim}
where {\tt epsabs} and {\tt epsrel} are used to set
the technical absolute and relative precision of the numerical integration.

\subsection{Initial parton momentum distributions}
\label{initialpdf}
The {\tt MarkovMC} library does not generate the initial parton momentum
distributions -- the user is supposed to provide the initial values of $x$- and
flavor-variables for the {\tt GenerateEvent} method. The actual generation of
the initial parton momentum distributions is therefore done by an external MC
application. We provide two examples of such an external environment. The first
one, in the folder {\tt Demo0} (see Section \ref{simpledemo}), simply uses 
fixed values of starting {\tt Xstart} and {\tt Flavor}. The second, a
more advanced example 
in the folder {\tt Demo1} (see Section \ref{Demo1}), uses the adaptive
MC generator {\tt TFoam} (part of the {\tt ROOT} system) to generate the initial
densities.
In {\tt Demo1}
we use the gluon ($G$) and quark singlet
($Q$) PDFs with three massless quarks in the following notation:
\begin{equation}
D_Q =\sum_i\Bigl( D_{q_i} + D_{\bar{q}_i} \Bigr)
\end{equation}
and
\begin{equation}
\label{inicond-quarks}
\begin{split}
D^0_{u}(x)&=  D^0_{u_{val}}(x)+\frac{1}{6} D^0_{sea}(x),\\
D^0_{d}(x)&=  D^0_{d_{val}}(x)+\frac{1}{6} D^0_{sea}(x),\\
D^0_{s}(x)&= 
D^0_{\bar{u}}(x)= 
D^0_{\bar{d}}(x)= 
D^0_{\bar{s}}(x)= \frac{1}{6} D^0_{sea}(x),\\
D^0_Q(x) &= D^0_{sea}(x)+D^0_{u_{val}}(x)+D^0_{d_{val}}(x).
\end{split}
\end{equation}
More details on the actual implementation of the {\tt TFoam}-based generation
can be found at the end of Section \ref{Demo1}.

\section{Installation instructions}
\label{sect:install}

This section instructs the user of {\tt EvolFMC v.2}
how to install the program on two system platforms:
{\tt Linux} and {\tt Mac~OS~X}.
The syntax of {\tt Linux} commands is given for the {\tt bash} shell.
The first few steps concern installation of the {\tt ROOT} package.
As explained earlier, the library {\tt MarkovMC} as such does not need {\tt ROOT}.
However, in the distributed version the {\tt ROOT} package is required
as a source of the random number generator for the library {\tt MarkovMC}.

The more advanced demonstration program {\tt Demo1}
exploits {\tt ROOT} as a histogramming package
and uses its persistency mechnism.

A step-by-step installation procedure of {\tt EvolFMC} looks as follows:
\begin{enumerate}
\item 
Check if {\tt ROOT} is installed in the system and find its location.
In the case there are several versions of {\tt ROOT} in the system,
choose the preferred one.
\item 
Check if the environmental variable {\tt ROOTSYS} is defined correctly:
{\tt echo \$ROOTSYS}.
If several versions of {\tt ROOT} are in the system,
define the {\tt ROOTSYS} variable as a path to the correct/preferred version:
\\ {\tt export ROOTSYS=path\_to\_your\_root}
\\
Check if the shell variable {\tt LD\_LIBRARY\_PATH} (or {\tt
DYLD\_LIBRARY\_PATH} in {\tt Mac~OS~X})
contains a correct path to the {\tt ROOT}'s library.
If not, then execute:
\\  under {\tt Linux:
\\  export LD\_LIBRARY\_PATH=\$LD\_LIBRARY\_PATH:\$ROOTSYS/lib}
\\  under {\tt Mac~OS~X:
\\  export DYLD\_LIBRARY\_PATH=\$DYLD\_LIBRARY\_PATH:\$ROOTSYS/lib}
\\ 
It is convenient to put this command  into the bash-shell configuration files:
{\tt .bashrc} or {\tt .bash\_profile}.
Finally, check if the {\tt \$PATH} variable includes a correct path to the {\tt ROOT} binaries.
\item
Add the path to the project to the variable
{\tt LD\_LIBRARY\_PATH} ({\tt DYLD\_LIBRARY\_PATH} in {\tt Mac~OS~X}).
This is done under {\tt Linux} by:{\tt
\\ export LD\_LIBRARY\_PATH=\$LD\_LIBRARY\_PATH:path\_to\_the\_project/lib}
\\ or under {\tt Mac~OS~X:
\\ export DYLD\_LIBRARY\_PATH=\$DYLD\_LIBRARY\_PATH:path\_to\_the\_project/lib}
\\ {\em Note:} For the demonstration programs {\tt Demo0} and {\tt Demo1} the user may skip the above commands 
because this path is already set in the appropriate {\tt Makefiles}.
\item 
Build the program from the commad line in the main project folder:
\\ {\tt autoreconf -i --force}
\\ under {\tt Linux:     ./configure}
\\ under {\tt MacOSX:    ./configure --enable-platform=macos}
\\ {\tt make}
\item
Test the correctness of the installation:
\\ {\tt (cd Demo0; ./verify\_benchmarks)}.
\\ For more details on the above test as well as on how to run two demonstration
  programs {\tt Demo0/Demo} and {\tt Demo1/Demo1Pr}
  see the next section.
\end{enumerate}

The authors have also managed {\tt EvolFMC} using two popular integrated
software development packages: {\tt Kdevelop} and {\tt Eclipse}.
Let us hint on how to initialize {\tt EvolFMC} 
as a project within these development tools:
\begin{enumerate}
\item
{\tt Kdevelop} 
\\ From the main folder of the project just type in the shell:
\\
{\tt kdevelop\&}
\\
   then from the menu 
      {\tt <project>}
   choose 
      {\tt <import>}
   and set 
      {\tt <project~type>} to {\tt <Generic~C++~Application~(Automake-based)>}.
   Next time you open {\tt Kdevelop}, the configuration files will be already in place.
\\
(Do not forget in the menu
     {\tt <project>$\to$<project~options>$\to$<configure~options>}:
     in the window {\tt <Configuration>} to choose
     {\em default} instead of {\em debug}%
     \footnote{Unless you really need debugging.}.)
\item
{\tt Eclipse}
\\ The configuration files are included,
   so it is enough to invoke {\tt Eclipse} and from the menu
   {\tt <File>$\to$<Import>$\to$<General>} choose {\tt <Existing~projects>}.
   The list of existing projects should appear, including the current {\tt EvolFMC}.
\end{enumerate}

\subsection{Testing correctness of installation}
Finally, let us explain how to test quickly the correctness of the installation.
This is done by means of executing a special ``benchmark test'' in form of the bash
script {\tt verify\_benchmarks}
included in the subfolder {\tt Demo0} of the distribution folder.
This test compiles and links the program {\tt Demo0/demo.cxx}, and then runs it
in a sequence for all eight implemented types of the evolution.
Text outputs from these runs,
containing a printout of the variable $x$, the flavor type and the MC weight for 100 events,
are produced and compared using the {\tt diff} utility
against the benchmark outputs stored by the authors of the code
in the subfolder {\tt Demo0/bmarks\_outputs}.
If the installation procedure and all the settings are correct,
then there should be no differences between
the stored and the current output disk files.
The stored outputs have been generated on the system iMac Core 2 duo with {\tt
gcc 4.2} under {\tt Ubuntu 8}.\\
Summarizing:
\begin{enumerate}
\item
 The benchmark test can be invoked as follows:
\\
{\tt cd Demo0}
\\
with the help of the {\tt bash} script
\\
  {\tt   ./verify\_benchmarks}
\\
\item
If needed, the user may create his/her own new set of benchmark output files
with the help of the bash-shell script in the {\tt Demo0} folder:
\\
  {\tt    ./create\_benchmarks}
\end{enumerate}
Alternatively, the above benchmark can be executed with the command
{\tt make bmark}.

\section{Two demonstration programs}
\label{sect:demos}
In the following we describe in a more detail two demonstration programs
in the subfolders $\tt Demo0$ and $\tt Demo1$,
which the user should run after installation of  {\tt EvolFMC}.
They are also meant as the templates for applications which use {\tt EvolFMC}
in studies related to the perturbative QCD -- most likely as a testing tool
for other MC programs implementing the QCD evolution of the parton momentum
distributions,
or as part of some bigger MC application.
These two programs are already built during installation,
see the previous section, and can be executed as follows:
\begin{enumerate}
\item
A simple demonstration program:
\\  {\tt cd Demo0}
\\  {\tt make start}
\item 
A more advanced demonstration program:
\\  {\tt cd Demo1/work}
\\  {\tt make start}
\\  Four histograms are recorded in the disk file.
    To visualize them:
\\  {\tt      make plot}  
\end{enumerate}
Let us describe these two demo programs
in a more detail.

\subsection{Simple demonstration program {\tt demo.cxx}}
\label{simpledemo}
The simple {\tt demo.cxx} program is included in the subfolder $\tt Demo0$ 
in order to demonstrate how to generate MC events 
for any of the eight evolution types supported by {\tt EvolFMC}.
The listing of the whole program {\tt demo.cxx} is given in the Appendix.
Let us present and explain the crucial instructions
in the {\tt demo.cxx} source code,
using the example of just one evolution type -- LO DGLAP:
\begin{verbatim}
[...]
#include "rndm.h"                                           (1)
#include "dglap_lo.h"
[...]
int main()
{	
  [...]
  rndm     * RNgen = new rndm();                            (2)
  dglap_lo * dglap_ll = new dglap_lo(T0,Lambda,
                 numberOfFlavors,RNgen,epsIRC,epsZmin);     (3)
  for(int i=0;i<numberOfEvents;i++)
  {
    dglap_ll->GenerateEvent(Tc,Xstart,weight,TStop,Flavor); (4)
    [...]
  }
  [...]
}
\end{verbatim}
\begin{enumerate}
\item
The appropriate header files are included:
one for the class of the random number generator
and another one for the class of the chosen evolution type.
\item
The object of the {\tt RNgen} class being the random number generator is created.
\item
Next, the MC generator object {\tt dglap\_ll} of the class {\tt dglap\_lo} is created.
It is the central object of the above code and it is used
to generate a series of the MC events.
The arguments of the {\tt dglap\_lo} constructor are:
the initial (starting) value of the evolution time {\tt Tmin},
the value of $\Lambda_{QCD}$ {\tt Lambda},
the number of active flavors {\tt numberOfFlavors}
and the pointer to the object of the random number generator {\tt RNgen}.
In addition, the parameter
{\tt epsIRC} is the technical cut-off used for regularizing
the distribution $1/(1-x)_+$ at $x=1$ in the kernel $\Keu^R_{ff'}$
of eq.\ (\ref{eq:evkernel}).
The final result will be independent of {\tt epsIRC},
if it is kept small enough.
The last parameter, {\tt epsZmin},
is the minimal requested value of the $x$ variable to be generated.
A non-zero value of the cut-off $\epsilon_{\min}$ is necessary only in the NLO cases,
due to the presence of the $(\ln z)/z$ singularity in the NLO DGLAP kernels.
$1/z$ in the kernels is cancelled for evolution of the momentum distributions
but $\ln z$ is still present,
hence some form of a cut-off is needed.
Note that the solution for $x>\epsilon_{\min}$ given by the program is always
independent of this cut-off, see Sect.\ \ref{smallparams} for more details%
\footnote{%
   On the contrary, the MC solution for $x<\epsilon_{\min}$ depends on this cut-off 
   and therefore should not be  trusted.}.
Note that both {\tt epsIRC} and {\tt epsZmin} have default values assigned by
the constructor and their redefinition, as done in the above example, is
optional.
\item
Finally, the MC events are generated with the method {\tt GenerateEvent}.
The meaning of the arguments of {\tt GenerateEvent} is the following:
\\
{\tt Tc} is a parameter for technical tests, not to be used.
The initial conditions of the evolution are set by
{\tt Xstart} and {\tt Flavor}, being the initial values of the $x$-variable
and the flavor type%
\footnote{
In general, {\tt Xstart} and {\tt Flavor}
will be generated according to some initial parton momentum distribution (see
{\tt Demo1})
-- here they are just set in the code.}.
{\tt TStop} is the maximal value of the evolution time,
{\tt weight} is the initial weight assigned to the event, normally set to 1,
{\tt epsSolver} is a technical parameter defining the accuracy of a procedure
for inverting numerically certain functions -- should not be modified!
The same parameters {\tt Xstart} and {\tt Flavor}
return the generated final value of the $x$-variable
and the final flavor type,
while {\tt weight} is the weight of the generated MC event.
A detailed history of the evolution is recorded inside
the object {\tt dglap\_lo}. In particular, the values of all generated
$x$- and flavor-variables, including their initial values, are stored there (in
the {\tt m\_x:MarkovMC} and {\tt m\_f:MarkovMC} matrices). All these
parameters can be accessed easily with the help of the 
functions {\tt GetParticle} and {\tt GetEventMultiplicity}.
\end{enumerate}
All other evolution types follow exactly the same pattern,
as can be seen in the {\tt demo.cxx} file.

The program {\tt demo.cxx} prints in the output the final $x$ and the final flavor
of the first 100 of the generated MC events.
After completing the MC generation it calculates and prints the average of the MC weight.
This average is equal to the sum over the final flavors integrated over the
final $x$-variable.
This, in turn, is {\em almost} equivalent to the unitary normalization
of the momentum distribution functions according to momentum sum rule.
It is {\em almost} equivalent because in the MC program
the lower limit $\epsilon_{\min}$  is imposed
on the value of the generated final $x$-variables, $x> \epsilon_{\min}$,
so there will be a tiny missing piece in the sum rule comming from the integral
from $0$ to $\epsilon_{\min}$.
Note that this integral contains only {\em integrable} singularities of the $\ln z$-type,
so by lowering $\epsilon_{\min}$ the sum rule can be tested to an arbitrary precision.

\subsection{More advanced demonstration program in folder {\tt Demo1}}
\label{Demo1}
The library of the classes in the folder {\tt MarkovMC}
is a collection of pure MC generators written in a clean and minimalistic way.
As shown in the previous simple demo,
it is easy to use the MC generator objects of these classes.
In the real life, the pure MC generators are only a small part
of a bigger code in which they are embedded.
Let us call this environmental code the {\em MC application} (MCAp)
and characterize briefly {\em functionality} and {\em data structures}
of such a MCAp.
The functionality of MCAp typically includes:
(a) running many times for various input data a MC event generator
and storing output data in a form of one and more dimensional histograms and/or
MC events, in the systems with one processor;
(b) the same in systems with many processors, many nodes (PC farms);
(c) visualization and quick analysis of the stored results after the MC ``production run''
is finished, or even while the MC production is still running,
in particular, 
(d) comparisons with the stored ``benchmark'' results 
    from the previous ``consolidated'' versions of the program,
(e) comparisons with the results of analytical and other non-Monte-Carlo 
    numerical calculations,
(f) comparisons between the stored MC results from the runs with different input
data and more.
The data structure of MCAp typically features:
(A) A collection (database) of the input parameters with clear distinction
between the parameters which are ``hardwired'' and changed only in very special
technical tests, the default paremeters which are rarely changed,
such that the user may normally ignore them,
and, finally, the important steering
parameters which are changed often or it is  even obligatory to (re-)define them.
(B) A data base of the output results from the consolidated well-tested
versions of the program, important results from long CPU time MC runs,
outputs used in the published works.
(C) Some degree of persistency mechanism for writing/reading structured
data in/from a disk file is absolutely necessary in organizing MCAp.
This persistency may be limited to data objects, such as MC events and histograms,
or include a possibility to write into a disk file the objects of the random number generator,
parts of the MC generator embedding important member data,
or even the complete MC event generator in the ``ready-to-go'' state.

In the presented distribution package we do not include the full scale
MCAp environment for the {\tt MarkovMC} library.
However,  the demonstration program in the folder {\tt Demo1} 
represents an essential step towards such an infrastructure.
The demonstration program {\tt Demo1/MainPr.cxx} features to a large extent
the points (a), (c), (d) and (C) of the above specification list
and is meant as an useful template for the further development.
Let us first overview it briefly, and more details will follow later on.

In this example the Markovian MC generator object {\tt m\_MMC},
being the instance of one of the eight classes of the {\tt MarkovMC} library,
is embedded as a member of the object {\tt m\_MCgen} of the container class {\tt MMCevol}.
The object {\tt m\_MMC} is created there, filled in with the input data,
and used to generate MC events using the methods
of the {\tt MMCevol} and {\tt MarkovianGen} classes.
Every MC event generated by {\tt m\_MCgen->m\_MMC}
is made available for histogramming.
The object {\tt m\_MCgen } is embedded (not created) in the object
of the class {\tt TRobolA}.
Methods of this class perform several functions:
book histograms, transfer input data and a
pointer of an external random number generator {\tt m\_RNG} object
into {\tt m\_MCgen},
generate a single MC event, fill-in histograms
and write histograms into a disk file.
The object of the {\tt TRobolA} class
does not contain, however, the main loop over the MC event generation,
nor creates the {\tt m\_RNG} and {\tt m\_MCgen} objects.
The main loop over the MC events is located and
managed in a rather special way in the main program {\tt MainPr},
while the objects of the external random number generator {\tt m\_RNG}
and of the MC generator {\tt m\_MCgen } are created in a separate
small script {\tt Demo1/work/Start.C}.
This scrips, run by {\tt ROOT} in the interpreter mode,
defines also all input data of the MC run.
Execution of {\tt Start.C}, building, running and stopping
the main program (as well as plotting the results) is managed by {\tt
Demo1/Makefile}.

The elements of the above five-level functionality and data structure
({\tt Makefile}, the main program, the {\tt TRobolA} class,
the {\tt MMCevol} class, the {\tt MarkovianGen} class/library),
sketched above, will be described in a more detail in the following.
At the first glance, this may look overcomplicated.
However, one has to remember that the fully functional MCAp,
defined in points (a)--(f), (A)--(C) above,
will unavoidably be rather complicated,
especially for running on the multinode PC farm.
The presented structure
exploits years of experience of the authors in developing many similar
MC infrastructures for running and testing MC event generators
\cite{koralw:1998, Jadach:2001mp, Jadach:2001uu, bhwide:1997,
bhlumi4:1996, Jadach:1999vf},
and we hope that it may be useful for others as a template to develop it further 
and/or customize to their own needs.

Let us now add more details on the {\tt Demo1} source code
and its execution.
All the {\tt C++} source code files specific to this demo program are located in
the folder
{\tt Demo1} while the input/output files can be found in its subfolder
{\tt Demo1/work}.
The {\tt main()} program is located in the file {\tt MainPr.cxx}.
It consists of three main parts corresponding to the initialization,
the generation of the MC event series
and the final part of the MC program.
The same three parts are present as three stages in the execution of the program.
In the first stage, the {\tt work/Start.C} script
creates three objects:
(1) an object of the {\tt MMCevol} class, that is the MC generator,
(2) an object of an external random number generator and
(3) an auxiliary object of the class {\tt TSemaf}.
These objects are initialized with the default data residing in the class constructors,
and then they are modified with the run-specific input data contained
explicitly in the code of the {\tt Start.C} script
(in particular, a random number seed may optionally be redefined at this point).
The object of {\tt TSemaf} contains data for administering
the main MC loop in the main program.
The above three objects are written by {\tt Start.C} into the disk files
{\em semaf.root} (the object {\tt TSemaf}) and {\em mcgen.root} (the objects
{\tt MMCevol} and random number generator).
In the above  and in the following the persistency
feature of the {\tt ROOT} system is exploited to facilitate
writing and reading the entire objects into/from the disk files.

The run-specific input data are all provided in {\tt work/Start.C},
including all steering parameters of the MC generator
the user wishes to reset from their  default values.
Examples of such settings together with comments,
specific to the {\tt Demo1} run, can be found inside {\tt work/Start.C}.

After executing {\tt Start.C}, the main program {\tt MainPr} comes
into action.
It reads all three stored objects from the disk files
{\em semaf.root} and {\em mcgen.root}
and creates a new object of the {\tt TRobolA} class containing
all histograms and the pointers to the MC generator {\tt m\_MCgen}
and the random number generator {\tt m\_RNG}.
The event generation loop is started.
The number of MC events is set by the variable {\tt NevTot}
which is read from the {\tt TSemaf} class object.
After generating every {\tt NGroup} events, the
partial results of the MC generation are stored in a disk
(the {\tt NGroup} variable is also taken from the {\tt TSemaf} object).
Before going to the next group of the event generation,
the status of the special semaphore-type flag in the {\tt TSemaf} object
in the disk file {\em semaf.root} is examined.
If the status flag is equal to {\tt "CONTINUE"}, then the event generation
is continued, otherwise, if it is equal to {\tt "STOP"},
then main loop and the MC generation is terminated.
The latter status flag can be reset by the user
interactively by calling {\tt make stop} from the subfolder
{\tt work}.
At the start of the program the status of this flag is
always set to {\tt "START"} and it is redefined to {\tt "CONTINUE"}
immediately after generating the first group of the MC events.
The class called {\tt TSemaf} is defined in the files {\tt (TSemaf.h, TSemaf.cxx)}.
The semaphore object contains also some auxiliary
information on the generated event sample, such as the number of events.
Once the event generation loop is terminated,
the programs enters into the short finalization
stage during which all the necessary statistics on the MC event sample are
calculated and all histograms of the {\tt TRobolA} class
are stored in the output file {\tt histo.root}.
Histograms are also recorded into {\tt histo.root} after generating
each group of MC events.

The source code of class {\tt TRobolA} is located in the
files  {\tt (TRobolA.h, TRobolA.cxx)}.
Its main task is to create and fill-in appropriate histograms.
This is done in three stages:
booking, filling and storing of the histograms with the
help of the class-member functions: 
{\tt  Hbook(),  Production(double\&), FileDump()}, respectively.
More details can be found in the source code.

The demonstration program {\tt Demo1/Demo1Pr} generates 
a sample of $10^5$ weighted events for the evolution of the parton momentum
 distributions from 
$Q_{\min}=1\,$GeV to $Q_{\max}=100\,$GeV.
In the course of the MC event generation, the program fills-in
histograms  of the gluon and quark-singlet parton momentum distributions
and the corresponding MC weights with the help of {\tt ROOT}.
These histograms are stored in the disk file {\tt histo.root}
written in the {\tt ROOT} format.
After completing the generation of the series of the MC events,
the stored histograms are plotted using the {\tt ROOT} graphics programs
and compared graphically with
the histograms  pre-computed and stored in the distribution folder.
The appropriate commands for running this demo program and
plotting the results are {\tt make start} and {\tt make plot}, see the beginning
of this section.

Another {\tt C++} program {\tt work/draw.cpp} is included
for analyzing/viewing the produced histograms and
comparing them with some pre-computed results. 
If the instalation of the MC program is done correctly,
the user-produced results should agree with the pre-computed
ones within statistical errors.
This program can be run in the interpreter mode of {\tt ROOT}
with the help of a single command {\tt make plot}.

Let us finally add a few details on embedding
the object {\tt m\_MMC} of the {\tt MarkovianGen} class
inside the ``wrapper'' object of the class {\tt MMCevol}.
The object {\tt m\_MMC} is created by {\tt MMCevol::Initialize()}
using the constructor of one of the eight classes implementing
a specific evolution type.
Which one to choose is decided by the steering data member
{\tt m\_EvolType} of the {\tt MMCevol} class.
Then, each MC event is generated by {\tt MMCevol::Generate()},
which calls {\tt m\_MMC->GenerateEvent()}.
Here, the method {\tt MarkovianGen::GenerateEvent()} does not have
any parameters (contrary to the one used in {\tt Demo0}).
This is why the getter {\tt MarkovianGen::SetTRange}
have to be used in {\tt MMCevol::Initialize} to define the range
of the evolution time.
Also, prior to {\tt m\_MMC->GenerateEvent()},
$x$ and the flavor of the initial parton are generated
using the {\tt TFoam} utility of {\tt ROOT}
and they are fed into the object {\tt m\_MMC} using the dedicated
setter {\tt MarkovianGen::SetInitParton}.
The corresponding part of the code in {\tt MMCevol::Generate()}
looks as follows:
\begin{verbatim}
// Generate primordial/initial parton
if( m_FoamI != NULL ){
  m_FoamI->MakeEvent();       // generate  x and parton type
  m_wt = m_FoamI->GetMCwt();  // get weight
} else
  m_wt =1.0;
/// Simulate QCD multiparton evolution
m_MMC->SetInitParton(m_flavIni,m_xIni); // Set initial parton
m_MMC->GenerateEvent();                 // Generate MC event
m_MMC->GetFinParton(m_Flav,m_X); // get final x and parton type
m_wt *= m_MMC->GetWt();          // combine MC weight
\end{verbatim}
The parton momentum distribution of the initial parton is provided to
the object {\tt m\_FoamI} of the {\tt TFoam} class
by the dedicated function {\tt MMCevol::Density} located in {\tt
Demo1/MMCevol.cxx} file. The user can easily modify here the shape of the
initial distributions.
The above organization is quite flexible and allows one to introduce
into the game more MC generator objects%
\footnote{%
Another possibility would be to inherit {\tt MMCevol} from the
{\tt MarkovianGen} class, but such a solution would probably be less flexible.},
more MC event objects, etc.

\section{Technical tests}
\label{sect:precision}
In this section we describe in some detail a variety of tests
that we have performed on the {\tt EvolFMC v.2} code
in order to verify its correctness and determine its overall technical precision.
To be specific, we have performed three different sets of
the technical comparisons of the code {\tt EvolFMC v.2}:
\\ (1) with the semianalytical code {\tt QCDNum16} \cite{qcdnum16}, 
\\ (2) with the semianalytical code {\tt APCheb40}
\cite{APCheb40,GolecBiernat:2007xv}, 
\\ (3) with the previous version of the {\tt EvolFMC} code: {\tt EvolFMC v.1},
and 
\\ (4) between different algorithms within the {\tt EvolFMC v.2}. 
\\
We will describe them in turn.
The target relative precision of the tests is $5\times 10^{-4}$ (half of a per mille).

At the end of this section we also present the weight distributions for
all the algorithms and we compare speed of the algorithms.

As the initial distributions at 1 GeV, for all of the tests we take
\begin{equation}
\label{inicond}
\begin{split}
D^0_G(x)&=1.908\cdot x^{-1.2}(1-x)^{5.0}, \\
D^0_{sea}(x)&=  0.6733\cdot x^{-1.2}(1-x)^{7.0}, \\
D^0_{u_{val}}(x)&=  2.187\cdot x^{-0.5}(1-x)^{3.0}, \\
D^0_{d_{val}}(x)&=  1.230\cdot x^{-0.5}(1-x)^{4.0}.
\end{split}
\end{equation}
The QCD constant $\Lambda_{0}=0.2457$ and $N_f=3$.
For each of the tests we use statistics of the order of $10^{10}$ MC points.

\subsection{Comparison with semianalytical code {\tt QCDNum16}}
The previous version of the code, {\tt
EvolFMC v.1}, has been tested against the {\tt QCDNum16} code
\cite{qcdnum16} for the standard DGLAP evolution. The demonstrated agreement
was
$5\times 10^{-4}$ for the LO case and $1\times 10^{-3}$ for the NLO case, see
\cite{GolecBiernat:2006xw}. For consistency we have repeated these comparisons
for {\tt EvolFMC v.2}.
\begin{figure}[!ht]
\hskip -8mm
\epsfig{file=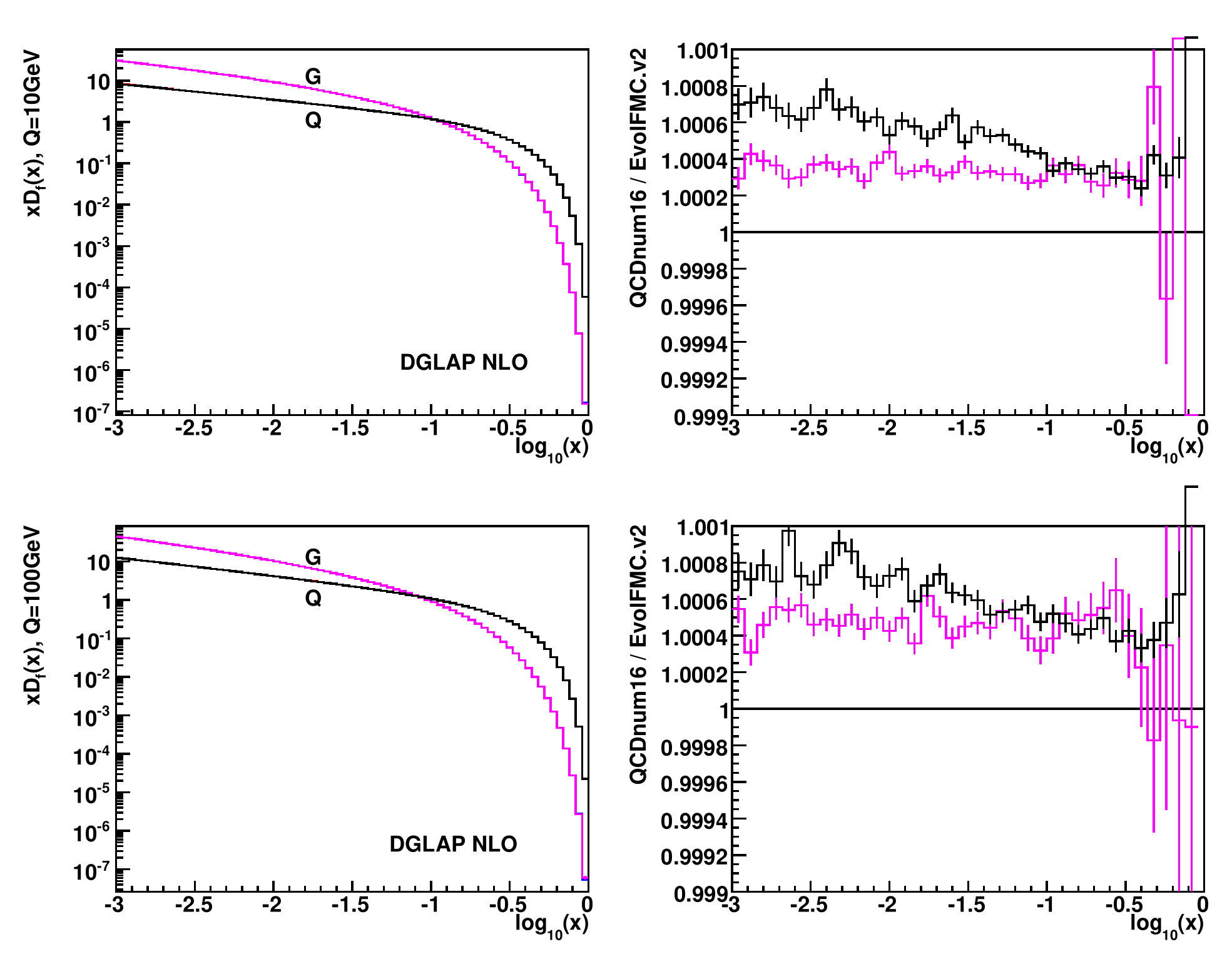, width=100mm,height=149mm,
angle=90 }
  \caption{\sf
{\underline{Left frames}}: the DGLAP evolutions
in the NLO approximation from {\tt QCDNum16} and  {\tt EvolFMC}
(the curves are indistinguishable). 
Upper curves (magenta and blue): the gluon $xD_G(x)$ distr.;
lower curves (black and red): the quark $xD_Q(x)$ distr.
{\underline{Right frames}}:
the ratio of {\tt QCDNum16} to {\tt EvolFMC} for the gluon  (magenta) and quark
(black)  
distributions.
{\underline{Top frames}}: the evolution up to 10 GeV.
{\underline{Bottom frames}}: the evolution up to 100 GeV.
}
  \label{fig:QCDNum16}
\end{figure}
In Fig.\ \ref{fig:QCDNum16} we show the DGLAP NLO evolution up to 
10 GeV and 100 GeV for both the gluon and quark singlet
$xD_f(x)$ distributions as well as the ratio of the {\tt QCDNum16} to {\tt
EvolFMC} results.
The {\tt QCDNum16} results are based on the extended grid size of 
$2000\times 600$. The agreement is of the order of $5\div 8\times 10^{-4}$.
We will demonstrate in the next subsection that these residual discrepancies are
to be attributed to {\tt QCDNum16}.
One has to remember that the comparisons with {\tt QCDNum16} are limited to the
standard DGLAP evolution only.

\subsection{Comparison with semianalytical code {\tt APCheb40}}
\begin{figure}[!ht]
\hskip -8mm
\epsfig{file=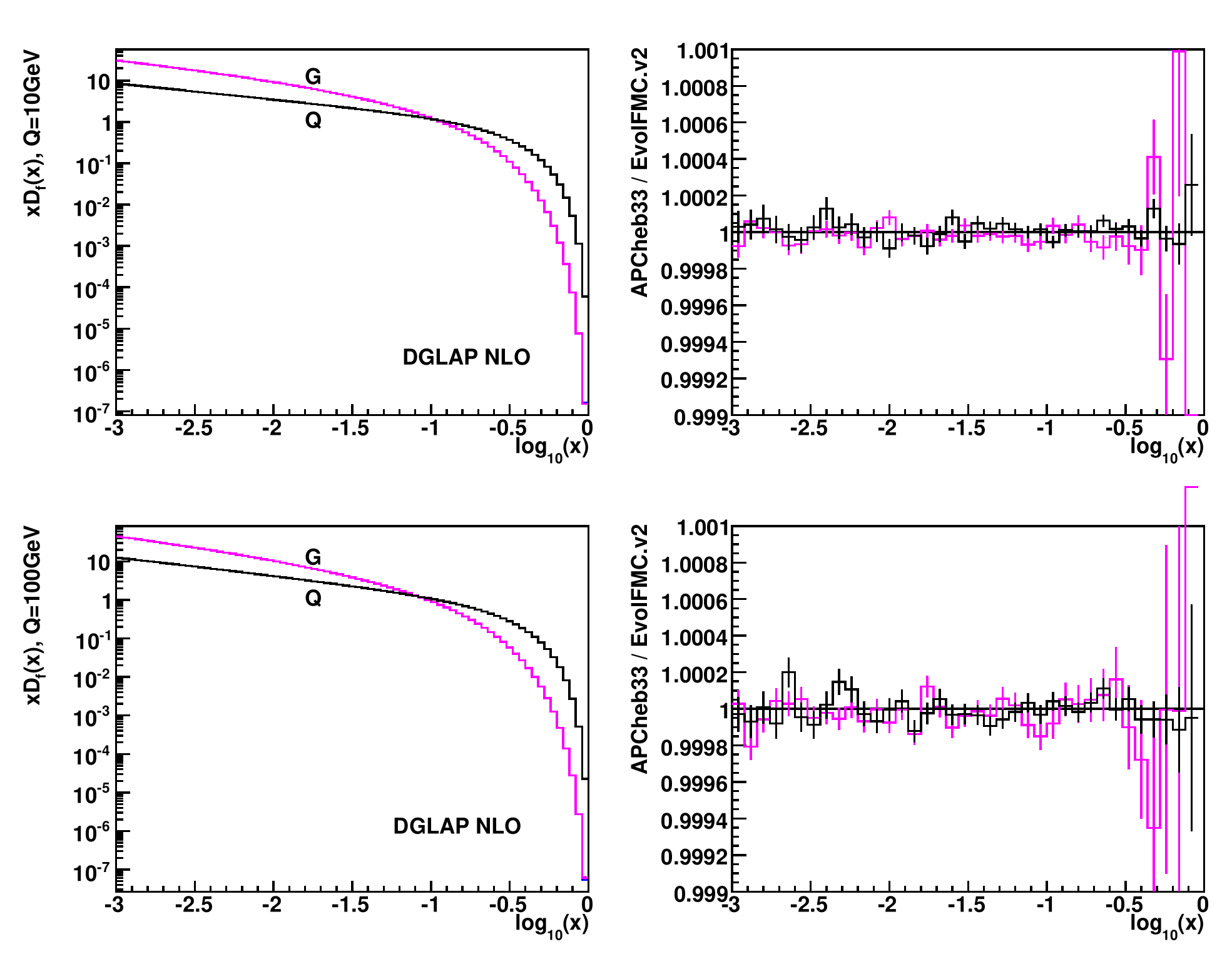, width=100mm,height=149mm,
angle=90 }
  \caption{\sf
{\underline{Left frames}}: the DGLAP evolutions
in the NLO approximation from {\tt APCheb33} and  {\tt EvolFMC}
(the curves are indistinguishable). 
Upper curves (magenta and blue): the gluon $xD_G(x)$ distr.;
lower curves (black and red): the quark $xD_Q(x)$ distr.
{\underline{Right frames}}:
the ratio of {\tt APCheb33} to {\tt EvolFMC} for the gluon  (magenta) and quark
(black)  
distributions.
{\underline{Top frames}}: the evolution up to 10 GeV.
{\underline{Bottom frames}}: the evolution up to 100 GeV.
}
  \label{fig:APCheb-nlo}
\end{figure}

It is the most independent test of the {\tt EvolFMC} code.
The semianalytical method of solving the evolution equations used by {\tt APCheb40}
is entirely different from the MC method and also the library of the
evolution kernels is completely independent. At first, in Fig.\
\ref{fig:APCheb-nlo}, we show the comparisons for the standard DGLAP NLO
evolution (for this comparison we used the older version, {\tt 33}, of the {\tt
APCheb} code).
We show the evolution up to 
10 GeV and 100 GeV for both the gluon and quark singlet
$xD_f(x)$ distributions as well as the ratio of the {\tt APCheb33} to {\tt
EvolFMC} results. As we see the agreement is remarkable, at the level of 
$1\times 10^{-4}$, limitted by the statistics. This result clarifies the source
of residual disagreement seen in the Fig.\ \ref{fig:QCDNum16}.
\begin{figure}[!ht]
\hskip -8mm
    \epsfig{file=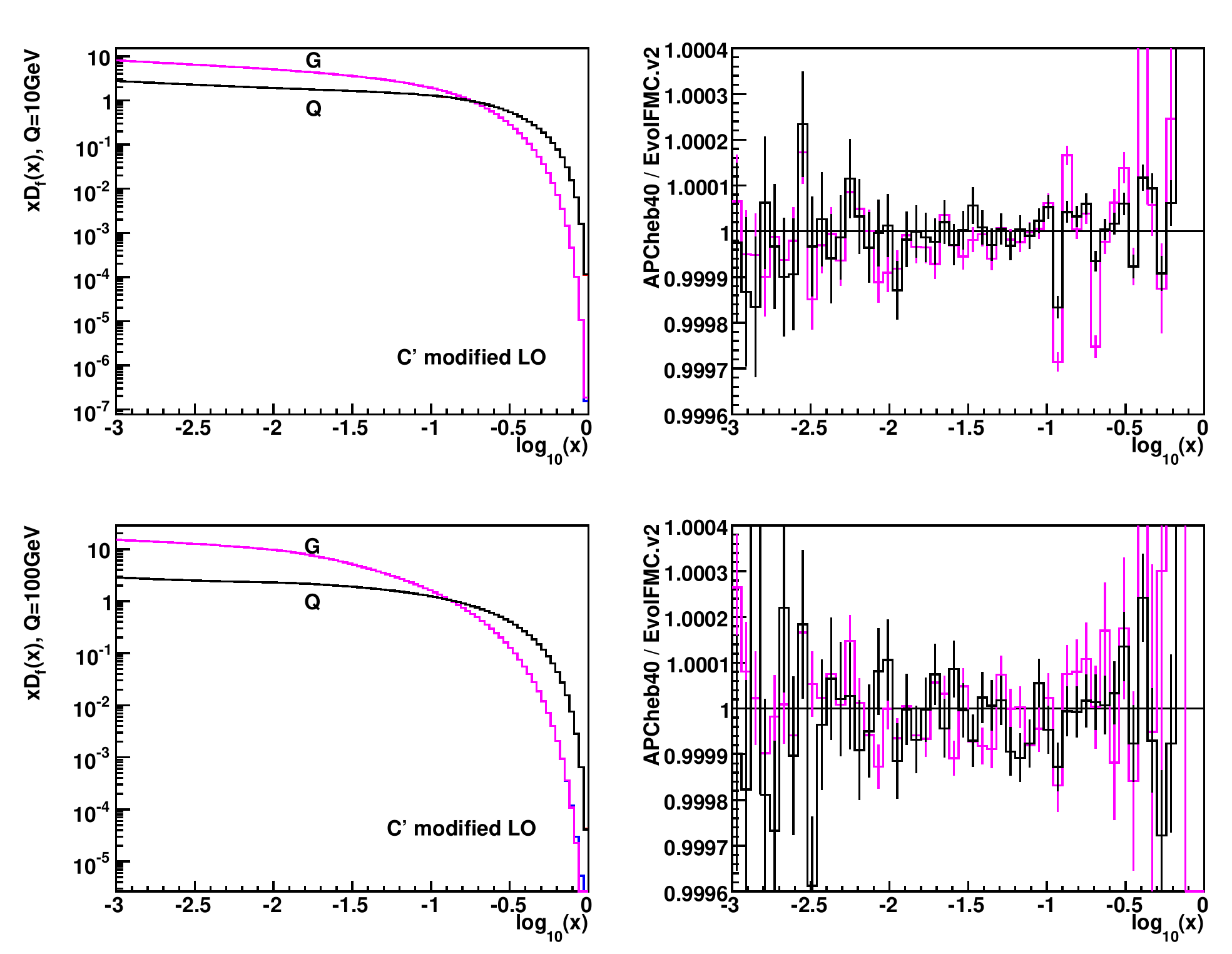,width=100mm,height=149mm,angle=90}
  \caption{\sf
{\underline{Left frames}}: the C'-type evolutions
in the modified LO approximation from {\tt APCheb40} and  {\tt EvolFMC}
(the curves are indistinguishable). 
Upper curves (magenta and blue): the gluon $xD_G(x)$ distr.;
lower curves (black and red): the quark $xD_Q(x)$ distr.
{\underline{Right frames}}:
the ratio of {\tt APCheb40} to {\tt EvolFMC} for the gluon  (magenta) and quark (black)  
distributions.
{\underline{Top frames}}: the evolution up to 10 GeV.
{\underline{Bottom frames}}: the evolution up to 100 GeV.
}
  \label{fig:APCheb10}
\end{figure}

The advantage of {\tt APCheb40}
over other semianalytical codes, such as {\tt QCDNum}~\cite{qcdnum16},
is that it has the option of the modified-DGLAP
evolution built-in, although only at the LO level, 
 see \cite{GolecBiernat:2007xv} for details. 
For the sake of comparisons we have modified the LO kernels in {\tt APCheb40}
in such a way that the coupling constant is implemented in the NLO approximation,
whereas the $z$-dependent parts remain in the LO approximation,
i.e.\ the modified kernels are
\begin{equation}
\label{kapc}
\begin{split}
&x\Keu^{R(B')\tt APCheb}_{f'f}(t,x,w)=
  \frac{\alpha_{NLO}(\ln(1-z)+t)}{2\pi}2zP_{f'f}^{R(0)}(z)
 \theta_{1-z>\lambda e^{-t}},
\\
&x\Keu^{R(C')\tt APCheb}_{f'f}(t,x,w)=
  \frac{\alpha_{NLO}(\ln(w-x)+t)}{2\pi}2zP_{f'f}^{R(0)}(z)
  \theta_{w-x>\lambda e^{-t}}.
\end{split}
\end{equation}
In Fig.\ \ref{fig:APCheb10} we show the C'-type evolution up to 
10 GeV and 
100 GeV for both the gluon and quark singlet
$xD_f(x)$ distributions as well as the ratio of the {\tt APCheb40} to {\tt EvolFMC} results.
The {\tt APCheb40} results are based on
the interpolation with 100 Chebyshev polynomials. The agreement is again
excellent, in most of the $x$-range below $1\times 10^{-4}$. 

To summarise,
the results of comparison with {\tt APCheb40} indicate that
technical precision of {\tt EvolFMC} is much better than our conservative
target of $5 \times 10^{-4}$.

\subsection{Comparison with {\tt EvolFMC} v.1}
\begin{figure}[!ht]
\hskip -6mm
    \epsfig{file=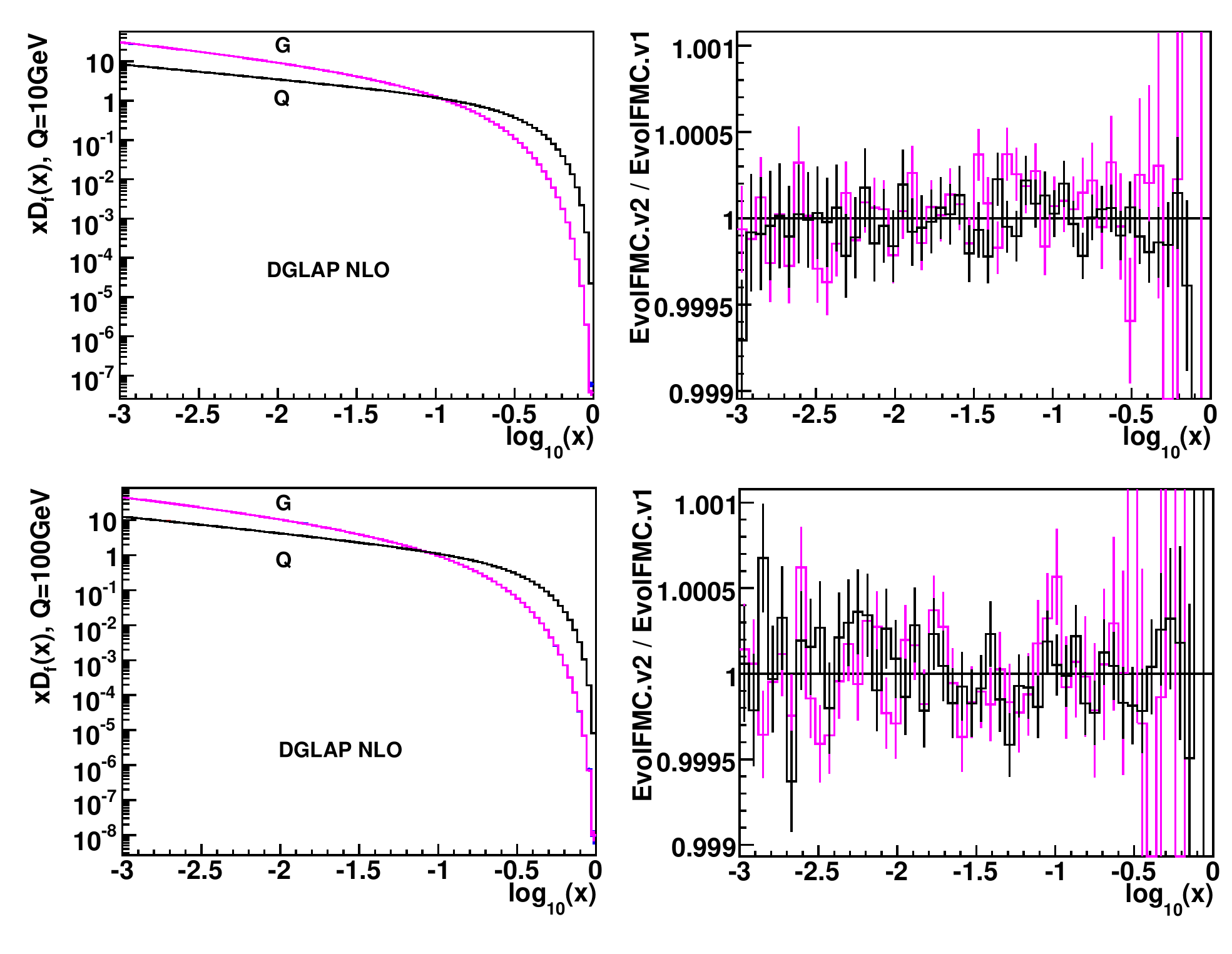,width=100mm,height=147mm,angle=90}
  \caption{\sf
{\underline{Left frames}}: the evolutions 
 from {\tt EvolFMC v.1} and  {\tt EvolFMC v.2} (the curves are indistinguishable) for
the DGLAP-type NLO evolutions.
Upper curves (magenta and blue): the gluon $xD_G(x)$ distr.;
lower curves (black and red): the quark $xD_Q(x)$ distr.
{\underline{Right frames}}:
the ratio of {\tt EvolFMC v.2} to {\tt EvolFMC v.1}
for the gluon (magenta) and quark (black)  distributions.
{\underline{Top frames}}: the evolution up to $Q=10$ GeV. 
{\underline{Bottom frames}}: the evolution up to $Q=100$ GeV. 
    }
  \label{fig:dglap_nlo_v1}
\end{figure}
\begin{figure}[!ht]
\hskip -6mm
    \epsfig{file=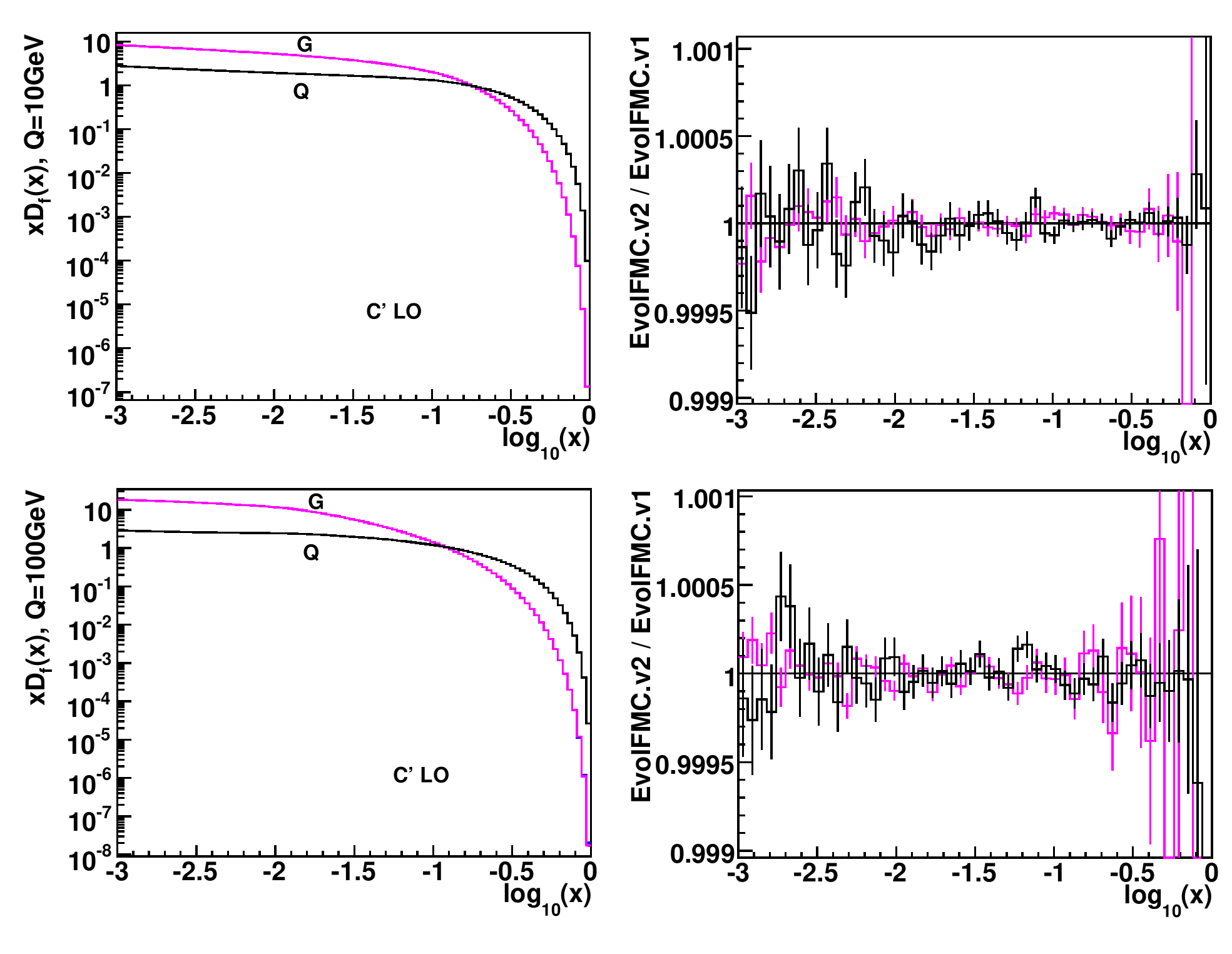,width=100mm,height=147mm,angle=90}
  \caption{\sf
{\underline{Left frames}}: the evolutions 
 from {\tt EvolFMC v.1} and  {\tt EvolFMC v.2} (the curves are indistinguishable) for
  modified-DGLAP C'-type LO evolutions.
Upper curves (magenta and blue): the gluon $xD_G(x)$ distr.;
lower curves (black and red): the quark $xD_Q(x)$ distr.
{\underline{Right frames}}:
the ratio of {\tt EvolFMC v.2} to {\tt EvolFMC v.1} for the gluon
(magenta) and quark (black)  distributions.
{\underline{Top frames}}: the evolution up to $Q=10$ GeV.
{\underline{Bottom frames}}: the evolution up to $Q=100$ GeV.
    }
  \label{fig:kt_lo_v1}
\end{figure}
The old {\tt EvolFMC} v.1 has been extensively tested both for the DGLAP (up to NLO) and
modified-DGLAP (LO only) cases with the overall relative precision tag of about $10^{-3}$ \cite{GolecBiernat:2006xw,GolecBiernat:2007xv}. 
As an example of the backward compatibility tests we show in Fig.\  \ref{fig:dglap_nlo_v1}
the comparison of {\tt EvolFMC} v.1 with {\tt EvolFMC} v.2 for 
the DGLAP-type NLO evolution   
and in Fig.\  \ref{fig:kt_lo_v1} for the modified-DGLAP C'-type LO evolution, both up to 10 and 100 GeV.
As we can see the results
agree within the statistical errors at the level of $5\times 10^{-4}$, as
desired. 

As we have explained in Introduction, the version {\tt v.1} of {\tt
EvolFMC} has different overall structure and organisation of the code as well as
different implementation of most of the methods. Therefore this comparison can
be regarded as a comparison with an ``almost'' independent code (for the
limited number of evolution types, of course).
\subsection{Comparison of different algorithms}
\begin{figure}[!ht]
\hskip -8mm
   \epsfig{file=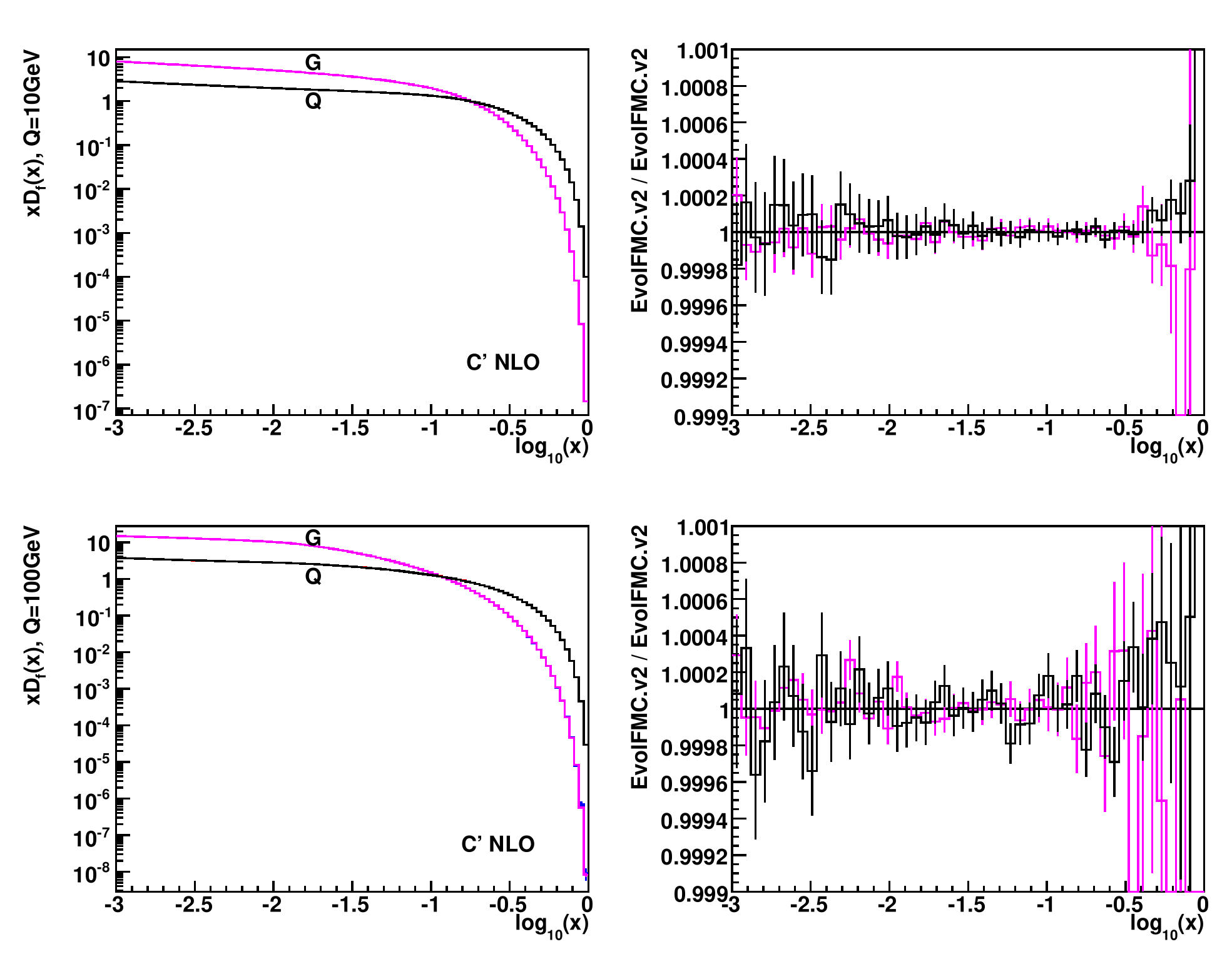,width=100mm,height=149mm,angle=90}
  \caption{\sf
{\underline{Left frames}}: the modified-DGLAP C'-type evolutions in NLO approximation 
from the two algorithms of  {\tt EvolFMC v.2} (the curves are indistinguishable). 
Upper curves (magenta and blue): the gluon $xD_G(x)$ distr.;
lower curves (black and red): the quark $xD_Q(x)$ distr.
{\underline{Right frames}}:
the ratio of the two algorithms of {\tt EvolFMC v.2} for the gluon
(magenta) and quark (black) distributions.
{\underline{Top frames}}:  the evolution up to 10 GeV.
{\underline{Bottom frames}}: the evolution up to 100 GeV.
    }
  \label{fig:kt_nlo_Q10}
\end{figure}
In this section we
show the tests of the most advanced evolution:  of the C'-type in the NLO approximation.
We have not found any other code that would solve this type of evolution.
For this reason we have implemented in the {\tt EvolFMC} code the second, auxiliary,
MC algorithm that solves this particular type of evolution.
The key difference of the auxilliary algorithm with respect to the main
algorithm is that the 
entire NLO correction (including modifications in the argument of the coupling
constant) is introduced as a weight and the algorithm itself is based on the LO
algorithm described in \cite{GolecBiernat:2007pu}, see
\cite{Stoklosa:2008nj} for details.

As an example in Fig.\ \ref{fig:kt_nlo_Q10} 
we show the comparisons of results of these two NLO algorithms of the C'-type
for two evolution time limits: 10 GeV and 100 GeV.
As one can see the agreement is well below the level of $5\times 10^{-4}$. This
result we consider as the principal test of the NLO C'-type evolution in the code
{\tt EvolFMC v.2}. Of course, the auxiliary algorithm shares with the previous ones a lot
of common parts of the code. These common parts however have been
extensively tested by all the comparisons described in previous subsections.

We conclude this series of tests with the conservative statement
that the {\tt EvolFMC v.2} code has the overall technical precision of $5\times10^{-4}$!

\subsection{Weight distributions and speed of algorithms}
\begin{figure}[!ht]
\hskip -8mm
   \epsfig{file=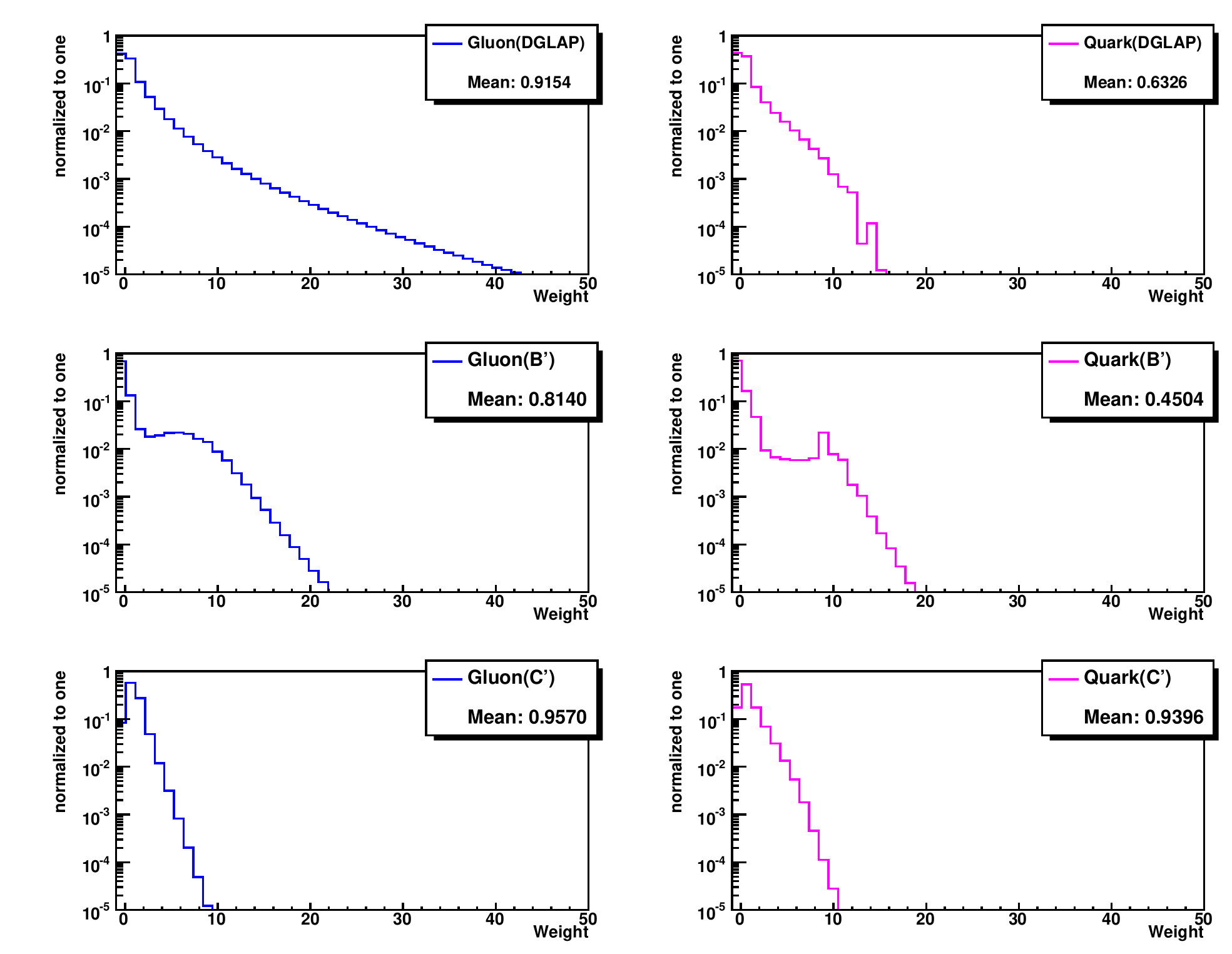,width=140mm,height=149mm,angle=90}
  \caption{\sf Weight distributions for NLO evolution to 100 GeV (normalized to
one).
{\underline{Left frames}}: gluon.
{\underline{Right frames}}: quark.
{\underline{Top frames}}:  DGLAP evolution.
{\underline{Central frames}}:  B'-type evolution.
{\underline{Bottom frames}}: C'-type evolution.
    }
  \label{fig:weights}
\end{figure}

In Fig.\ \ref{fig:weights} we show the weight distributions for all three main
algorithms included in the program {\tt EvolFMC v.2}. The evolutions are of the
NLO type up to 100 GeV.
The weights are well behaved and fall down exponentially. The
distributions of modified evolutions have shorter tails as compared to the DGLAP
case. This is the consequence of the finite IR cut-off used in the modified
kernels, as compared to the infinitesimal one used in the DGLAP case. The plots
indicate that the conversion to the unweighted events should be quite
efficient, especially in the case of moderate target precision level. One has
to remember that in the current version of the program we did not perform
any additional weight optimization, as we restricted ourselves to the weighted
events only. Such an optimization could reduce the tails of the weight
distributions even further. 

\begin{figure}[!ht]
\hskip -8mm
  
\epsfig{file=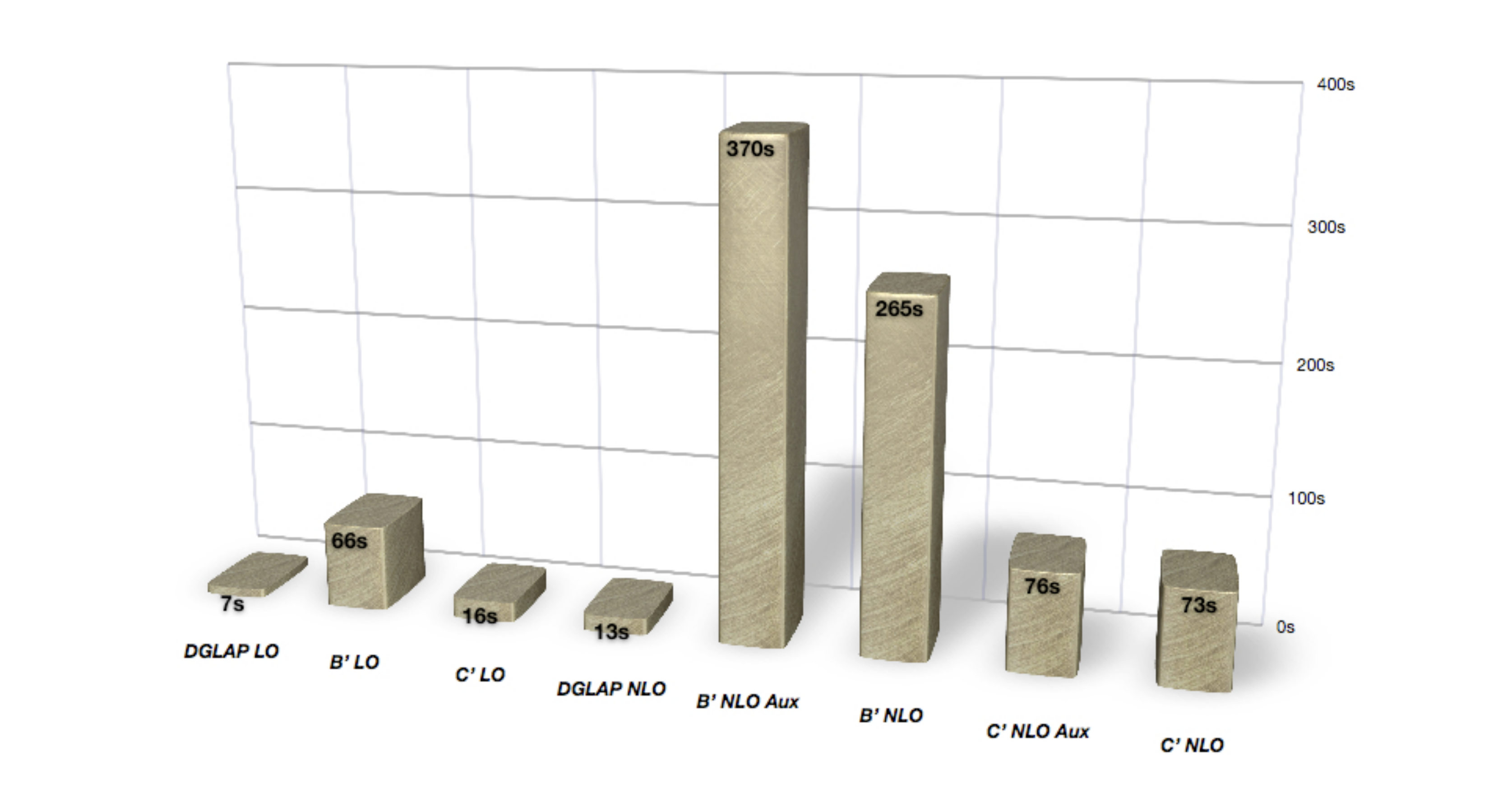,width=140mm,height=80mm}
  \caption{\sf Comparison of the speed of all five algorithms. Plotted
times (in seconds) refer to generation of $10^6$ events and evolution to 100
GeV. First three blocks illustrate the LO evolution (DGLAP, B' and C', respectively), 
following five blocks the NLO case (DGLAP, B'-auxilary, B'-main, C'-auxiliary 
and C'-main, respectively).
    }
  \label{fig:speed}
\end{figure}

Finally, in Fig.\ \ref{fig:speed} we compare speed of all five algorithms
included in the program {\tt EvolFMC v.2} both in LO and NLO cases. As we see,
despite their complexity and the presence of internal one-dimensional numerical
integration, the NLO algorithms are slower only by a factor of few as compared
to the very fast LO ones. The slowest one is the B'-type algorithm. Let us note
that the algorithms are only partly optimized with respect to speed.

\section{Summary}
\label{sect:summary}
In this paper we have presented the program {\tt EvolFMC v.2}.
It is a Monte Carlo generator that solves some of the QCD evolution
equations for the parton momentum distributions in the weighted event mode.
In the current version we have implemented the DGLAP evolution in the LO and NLO
approximations and two modified-DGLAP type evolutions.
These modifications include the replacement of the argument of the coupling constant
by a more complicated functions of the $x$ and $t$ variables.
In one case (C') this function reconstructs the transverse momentum of the emitted partons
and in the other case (B') the approximate transverse momentum.
Both of these modified-DGLAP evolutions are implemented in the LO as well as the NLO approximation.
The code is written in the {\tt C++} language and has the modular structure,
easy to extend to additional evolution types.
The main limitation of the code is the fact that quarks are treated as massless.
We have studied the technical precision of the program by means of extensive comparisons
with the non-MC numerical program {\tt APCheb40}, with the previous version of the code, {\tt EvolFMC v.1},
and by comparing two independent algorithms for the same evolution implemented in {\tt EvolFMC v.2}.
These tests have proved that the technical precision of the {\tt EvolFMC v.2} is at least $5\times 10^{-4}$.
To our knowledge it is the first so-precise solution of the QCD evolution equations
by means of the Monte Carlo methods.

\section{Acknowledgements}
Authors would like to thank K. Golec-Biernat and Z. W\c{a}s for discussions and comments.

\section{Appendix}

In this Appendix we list the source code of the simple {\tt demo.cxx} program from the {\tt Demo0}
folder.

{\small
\begin{verbatim}

#include<stdio.h>
#include<math.h>

#include "rndm.h"

#include "bprim_nlo.h"
#include "bprim_lo.h"
#include "bprim_nlo_aux.h"
#include "cprim_lo.h"
#include "cprim_nlo.h"
#include "cprim_nlo_aux.h"
#include "dglap_lo.h"
#include "dglap_nlo.h"

#define DGLAP_LO      0
#define DGLAP_NLO     1
#define CPRIM_LO      2
#define CPRIM_NLO     3
#define CPRIM_NLO_AUX 4
#define BPRIM_LO      5
#define BPRIM_NLO     6
#define BPRIM_NLO_AUX 7

double Lambda         = 0.25;
double numberOfFlavor = 3;
double Qmin           = 1;
double Qmax           = 100;
int TypeOfGenerator   = 0;

rndm * RNgen;
int numberOfEvents = 100;

int main(int argc, char *argv[])
{
	if (argc>1)
		TypeOfGenerator = (atoi(*(++argv))<8) ? atoi(*argv) : 0;

  RNgen = new rndm();
  double T0 = log(Qmin);
  dglap_lo  * dglap_ll
               = new dglap_lo(     T0,Lambda,numberOfFlavor,RNgen);
  dglap_nlo * dglap_nll
               = new dglap_nlo(    T0,Lambda,numberOfFlavor,RNgen);
  CPrim_lo  * cprim_ll
               = new CPrim_lo(     T0,Lambda,numberOfFlavor,RNgen);
  CPrim_nlo * cprim_nll 
               = new CPrim_nlo(    T0,Lambda,numberOfFlavor,RNgen);
  CPrim_nlo_aux * cprim_nll_aux 
               = new CPrim_nlo_aux(T0,Lambda,numberOfFlavor,RNgen);
  BPrim_nlo * bprim_nlo
               = new BPrim_nlo(    T0,Lambda,numberOfFlavor,RNgen);
  BPrim_lo  * bprim_lo 
               = new BPrim_lo(     T0,Lambda,numberOfFlavor,RNgen);
  BPrim_nlo_aux * bprim_nlo_aux 
               = new BPrim_nlo_aux(T0,Lambda,numberOfFlavor,RNgen);

	dglap_ll->printLogo();

	int init_Flavor = 0;
	double init_Xstart = 1.0;

printf("*******************Starting parameters*****************\n");
	printf("Qmin=%4.2f\n", Qmin);
	printf("Qmax=%4.2f\n", Qmax);
	printf("Type of Equation: %d\n", TypeOfGenerator);
	printf("QCDLambda=%2.4f\n", Lambda);
	printf("numberOfFlavors=%d\n", numberOfFlavor);
	printf("Events to generate:%d\n", numberOfEvents);

printf("*******************Initial Conditions******************\n");
	printf("Flavor: %d\n", init_Flavor);
	printf("Xstart: %1.2f\n", init_Xstart);
printf("*************************Events************************\n");

	double TStop = log(Qmax);
	double Tc;

	double sumwt = 0.0;
	double sumwt2= 0.0;
	int NoEvents =0;

	for (int i=0; i<numberOfEvents; i++)
	{
		int Flavor = init_Flavor;
		double Xstart = init_Xstart;

		double weight = 1.0;

		switch (TypeOfGenerator)
		{
		case CPRIM_NLO:
		        cprim_nll->GenerateEvent(Tc, Xstart, weight, TStop, Flavor);
		        break;
			case CPRIM_NLO_AUX:
		        cprim_nll_aux->GenerateEvent(Tc, Xstart, weight, TStop, Flavor);
		        break;
		case CPRIM_LO:
		        cprim_ll->GenerateEvent(Tc, Xstart, weight, TStop, Flavor);
		        break;
		case DGLAP_LO:
		        dglap_ll->GenerateEvent(Tc, Xstart, weight, TStop, Flavor);
		        break;
			case DGLAP_NLO:
		        dglap_nll->GenerateEvent(Tc, Xstart, weight, TStop, Flavor);
		        break;
		case BPRIM_LO:
		        bprim_lo->GenerateEvent(Tc, Xstart, weight, TStop, Flavor);
		        break;
		case BPRIM_NLO:
		        bprim_nlo->GenerateEvent(Tc, Xstart, weight, TStop, Flavor);
		        break;
		case BPRIM_NLO_AUX:
		        bprim_nlo_aux->GenerateEvent(Tc, Xstart, weight, TStop, Flavor);
		        break;
		}
		sumwt =sumwt + weight;
		sumwt2=sumwt2 + weight*weight;
		NoEvents = NoEvents+1;
		if (NoEvents<100)
		{
			printf("X=%1.6f, weight=%1.6f, flavor=%d\n", Xstart, weight, Flavor);
		}
	}
double integral=sumwt/NoEvents;
double error = sqrt( (sumwt2/NoEvents 
                   - (sumwt/NoEvents)*(sumwt/NoEvents)) /NoEvents);

printf("***********************Output**************************\n");
printf("\n [sum_f=flavor] [int_eps^1 dx] D_f(x,t)= %1.6f +- %1.6f\n\n"
                                                  ,integral, error);
printf("*******************************************************\n");

return 0;
}
\end{verbatim} }




\end{document}